\documentclass[twocolumn]{aastex701}

\usepackage{amsmath}
\usepackage{booktabs}
\usepackage[version=4]{mhchem}

\newcommand{\onedlogz}{2.23^{+0.07}_{-0.07}}
\newcommand{\onedlinearz}{170^{+30}_{-24}}
\newcommand{\twodlogz}{1.11^{+0.25}_{-0.20}}
\newcommand{\twodlinearz}{13^{+10}_{-5}}

\newcommand{\prtonedlinearz}{112^{+18}_{-16}}

\newcommand{\prtlinearz}{6.8^{+3.6}_{-2.0}}
\newcommand{\onedhostz}{73^{+15}_{-12}}
\newcommand{\twodhostz}{5.6^{+4.6}_{-2.2}}
\newcommand{\wasptrendz}{5.7}
\newcommand{\onedtrendratio}{12.7}
\newcommand{\combineddeltaevidence}{11.84}

\newcommand{\combinedbayesfactor}{1.4\times10^{5}}

\newcommand{\twodchisq}{479.1}

\newcommand{\prtchisq}{482.8}

\newcommand{\limbdeltascatt}{3.3}
\newcommand{\limbweight}{0.49^{+0.04}_{-0.05}}

\newcommand{\metallicityratio}{13.1^{+8.4}_{-5.9}}

\newcommand{\eveninglimbredchisq}{1.10}
\newcommand{\morninglimbredchisq}{0.79}

\newcommand{\jointlimbredchisq}{0.94}

\newcommand{\platonsevenvalidationdraws}{1000}
\newcommand{\platonsevenmaxppm}{0.0066}
\newcommand{\platonsevenrmsppm}{0.0022}
\newcommand{\platonsevenmaxsigma}{3\times10^{-4}}
\newcommand{\platonsevenmetalshift}{3.08\times10^{-5}}
\newcommand{\platonsevenessfraction}{1\times10^{-4}}

\newcommand{\platon}{\texttt{PLATON}}

\newcommand{\zsun}{Z_{\rm atm}/Z_\odot}
\shorttitle{Two Limbs from a Limb-averaged  Spectrum}
\shortauthors{Fairnington et al.}

\defcitealias{Mukherjee_2026a}{M26}
\defcitealias{Ahrer:25a}{A25}
\begin{document}

\title{Recovering the Morning and Evening Limbs of WASP-94Ab from Its Limb-Averaged JWST Transmission Spectrum}

\correspondingauthor{Tyler R. Fairnington}
\email{tfairnington@uchicago.edu}

\author[0000-0002-0692-7822]{Tyler R. Fairnington}
\affiliation{Department of Astronomy \& Astrophysics, University of Chicago, 5640 South Ellis Avenue, Chicago, IL 60637, USA}
\email{tfairnington@uchicago.edu}

\author[0000-0002-3328-1203]{Michael Radica}
\altaffiliation{NSERC Postdoctoral Fellow}
\affiliation{Department of Astronomy \& Astrophysics, University of Chicago, 5640 South Ellis Avenue, Chicago, IL 60637, USA}
\email{radicamc@uchicago.edu}

\author[0000-0002-0659-1783]{Michael Zhang}
\affiliation{Department of Astronomy \& Astrophysics, University of Chicago, 5640 South Ellis Avenue, Chicago, IL 60637, USA}
\email{zmzhang@uchicago.edu}

\author[0000-0002-1337-9051]{Eliza M.-R. Kempton}
\affiliation{Department of Astronomy \& Astrophysics, University of Chicago, 5640 South Ellis Avenue, Chicago, IL 60637, USA}
\email{ekempton@uchicago.edu}

\author[0000-0003-4733-6532]{Jacob L. Bean}
\affiliation{Department of Astronomy \& Astrophysics, University of Chicago, 5640 South Ellis Avenue, Chicago, IL 60637, USA}
\email{jacobbean@uchicago.edu}

\author[0000-0003-0973-8426]{Eva-Maria Ahrer}
\affiliation{Max Planck Institute for Astronomy (MPIA), K\"onigstuhl 17, 69117 Heidelberg, Germany}
\email{ahrer@mpia.de}

\author[0000-0002-2454-768X]{Arjun B. Savel}
\affiliation{Astronomy Department, University of Maryland, College Park, 4296 Stadium Dr., College Park, MD 20742 USA}
\affiliation{Department of Astronomy, California Institute of Technology, 1200 E. California Blvd., Pasadena, CA 91125, USA}
\email{savel@caltech.edu}

\author[0000-0001-9358-1841]{Pascal S. D'Andrea}
\affiliation{Institute for Particle Physics and Astrophysics, ETH Zurich, CH-8093 Zurich, Switzerland}
\affiliation{Department of Astronomy \& Astrophysics, University of Chicago, 5640 South Ellis Avenue, Chicago, IL 60637, USA}\email{pdandrea@ethz.ch}

\begin{abstract}
Recent JWST observations have revealed that the morning and evening limbs of the hot Jupiter WASP-94Ab have very different transmission spectra. An analysis of these resolved limb spectra yielded a metallicity more than an order of magnitude below that obtained from the limb-averaged spectrum, which calls into question the accuracy of traditional retrievals for heterogenous planets. Here we analyze the limb-averaged JWST NIRISS and NIRSpec transmission spectra of WASP-94Ab using both a standard one-temperature-profile (1-TP) retrieval with patchy clouds, and a new two-temperature-profile (2-TP) retrieval. The 2-TP retrieval assumes a linear combination of two atmospheric components that share the same bulk elemental composition but have independent thermal structures, chemistries, and aerosol properties. This model is implemented in \texttt{PLATON 7.0}, a \texttt{JAX}-accelerated code that is over 100 times faster than its predecessor, which is key to making the retrieval computationally tractable.
The 2-TP model is favored over the 1-TP model ($K\simeq\combinedbayesfactor$) and yields $\zsun=\twodlinearz$, compared with $\onedlinearz$ for the 1-TP model. This is broadly consistent with the $2.9_{-1.6}^{+3.7}\times$ solar that \citet{Mukherjee_2026a} derived from NIRISS/SOSS limb spectra alone. The remaining offset likely reflects the additional constraints on carbon-bearing species from our NIRSpec/G395H data.
The 2-TP retrieval  predicts two, equally weighted components that reproduce the limb-resolved spectra across both instruments without prior knowledge of them. 
This work demonstrates that compositional information survives limb averaging for WASP-94Ab, suggesting that 2-TP retrievals can be used to identify limb-asymmetry and extend metallicity studies beyond the small set of planets amenable to limb-resolved spectroscopy.
\end{abstract}


\section{Introduction}\label{sec:intro}

Most transmission spectroscopy retrievals fit the spectrum of an exoplanet's terminator with one atmospheric column \citep{Madhu:18}. Yet general circulation models of tidally locked giant planets predict substantial temperature differences and circulation across their atmospheres \citep[e.g.,][]{Showman:09, Rauscher:12, Mayne:17}.  Simulations that include cloud formation additionally predict spatial variations in aerosol opacity \citep{Parmentier:16, Lines:18, Powell:19, roman_clouds_2021}, while coupled chemical and dynamical calculations demonstrate composition gradients and transport-driven disequilibrium \citep{Mendoca:18, zamyatina_quenching-driven_2024}.
These gradients in atmospheric properties are predicted to occur along multiple spatial axes, e.g., from pole to equator, day to night, or east to west. The latter, in particular, has come to the fore in the JWST era, with multiple planets showing clear signatures of east-west (or ``morning-evening'') spectral differences through specialized time-resolved transit analyses \citep{espinoza_inhomogeneous_2024, murphy_evidence_2024, Ahrer:25a, fu_overcast_2025, Mukherjee_2026a}.

If significant spatial inhomogeneities exist, a single thermal profile, composition, and aerosol prescription cannot reliably represent the entire terminator. Standard 1-D single profile (1-TP) retrievals can consequently return biased atmospheric properties despite statistically acceptable fits, as multiple forward modeling studies showed in the lead-up to JWST \citep[e.g.,][]{Line:16, Caldas:19, MacDonald:20, Nixon:22, Pluriel:22, Welbanks:22}. The remedies to this 1-D biasing problem proposed by the aforementioned studies have not yet been widely applied, likely due to the significant computational cost of retrieving simultaneously on a multi-component terminator. A notable exception is the \citet{Line:16} patchy-cloud formulation, which is commonly used to fit for a simplified two-component atmosphere with clear and cloudy regions that share global temperature--pressure (TP) and abundance profiles.

Instead of more complex retrievals, the emerging gold standard with JWST has been to make use of the shape of transit light curves to spatially resolve the distinct morning and evening terminators \citep[e.g.,][]{espinoza_inhomogeneous_2024, murphy_evidence_2024, murphy_panchromatic_2025, ahrer_tracing_2025, fu_overcast_2025, Mukherjee_2026a}. The hot Jupiter WASP-94Ab provides a particularly poignant example of the benefits of this approach. From a limb-resolved analysis of NIRISS/SOSS data, \citet[][hereafter \citetalias{Mukherjee_2026a}]{Mukherjee_2026a} inferred a colder morning limb almost completely muted by aerosols and a hotter evening limb rich in molecular features. Retrievals performed on the individual planetary limbs clearly demonstrate the 1-TP bias: their evening-limb retrieval returned a metallicity roughly 1.6\,dex lower than that inferred from their traditional 1-TP retrieval of the limb-averaged spectrum.

The \citetalias{Mukherjee_2026a} study poses a major challenge for transmission spectroscopy studies of exoplanet atmospheres. If the correct atmospheric metallicity can only be recovered from limb-resolved analyses for WASP-94Ab, then this is likely also true for other planets as well. Yet only a small subset of the highest S/N targets are amenable to such analyses because they require high precision sampling of the brief ingress and egress transit phases.  
A potential solution is found in the work of \citet{Chen:25} who demonstrated that for a different planet, WASP-39b, the limb-averaged data retain distinct limb signals matching phase-resolved observations when a two-component retrieval is performed \citep{MacDonald:20, Welbanks:22}. 

In what follows, we extend the work of \citet{Chen:25} and show that we are able to directly recover WASP-94Ab's morning and evening limb spectra from its limb-averaged spectrum. Our ``two-temperature-profile'' (2-TP) retrieval approach allows us to fit independently for the aerosol and thermal structure on two disparate regions of the planet, while enforcing a global set of underlying elemental abundances and deep atmosphere conditions.  This highly multi-dimensional retrieval is enabled by a new \texttt{JAX}-optimized implementation of the \texttt{PLATON} retrieval code that is more than two orders of magnitude faster than its legacy implementation \citep{PLATON7}.

\section{Observations and Data Analysis}\label{sec:data}

\subsection{Observations and Reduction}

We measured the limb-averaged transmission spectrum of WASP-94Ab by reanalyzing the NIRSpec/G395H transit observed on 2024 June 7 as part of GO~3154 (PI: E.-M.~Ahrer) and the NIRISS/SOSS transit observed on 2024 October 11 in GO~5924 (PI: D.~Sing). These observations were originally published in \citet[][hereafter \citetalias{Ahrer:25a}]{Ahrer:25a} and \citetalias{Mukherjee_2026a}, respectively.

We re-reduced the uncalibrated detector products with \texttt{exoTEDRF} \citep{radica_awesome_2023, feinstein_early_2023, Radica:24}, closely following the procedures of \citet{radica_supersolar_2026}. Crucially, instead of manually tuning many of the reduction parameters, we make use of an automated coordinate-descent optimization wrapper for \texttt{exoTEDRF}. The optimizer will be elaborated upon in full detail in Fairnington et al.\ (in prep.), but in short, it proceeds sequentially though reduction parameters, varying each in turn over a user-defined grid to determine the values that minimize the point-to-point scatter of the spectroscopic light curves. 
For the optimizable parameters, we select the jump detection threshold, trace and background mask widths during 1/$f$-correction, spatial and temporal outlier rejection thresholds during the bad pixel interpolation, and the extraction aperture width. The NIRSpec NRS1 and NRS2 detectors are optimized independently. We use the \texttt{scale-achromatic} and \texttt{median} methods for the correction of 1/$f$ noise in the NIRISS and NIRSpec observations respectively, and the time-domain cosmic ray-flagging routine \citep{radica_muted_2024}. We perform a box extraction with an aperture width of 32 pixels for NIRISS and 8 pixels for NIRSpec.

Several zeroth-order sources and dispersed field-star traces contaminate the NIRISS detector. We use the accompanying F277W exposure to identify the zeroth-order sources \citep[e.g.,][]{radica_awesome_2023, fournier-tondreau_near-infrared_2024} and mask the affected regions following \citetalias{Mukherjee_2026a}. Only a narrow interval of the second SOSS order remains uncontaminated, so we exclude that order altogether. The retained SOSS first order and G395H spectra span approximately $0.85$--$5.0\,\mu{\rm m}$.

\subsection{Light Curve Analysis}

For each dataset, we first construct white-light curves to constrain the transit geometry. We model the planet with \texttt{jaxoplanet} \citep{jaxoplanet} and adopt the power-2 limb-darkening law, as it has been shown to produce superior fit quality over all other two-parameter laws, including the typically used quadratic variant  \citep{power2,power_2_better}. Priors on the limb-darkening coefficients are generated with \texttt{ExoTiC-LD} \citep{exotic-LD} and the three-dimensional Stagger stellar-atmosphere grid \citep{stagger}. We sample $T_{\rm eff}$, $\log g_\star$, and $[\mathrm{Fe/H}]_\star$ over their $5\sigma$ ranges and likelihood-weight the resulting coefficient distributions, thereby propagating stellar-parameter uncertainties into the transit depths.

The white-light fits include the mid-transit time $t_0$, impact parameter $b$, total duration $T_{14}$, radius ratio $R_{\rm p}/R_\star$, two limb-darkening coefficients, the transit zero point, a linear trend with time, as well as a jitter term added in quadrature to the flux errors. We fix the orbital period to 3.9502001 days \citep{Kokori:23}. The spectroscopic fits hold $(t_0,b,T_{14})$ at the white-light posterior medians and independently infer $R_{\rm p}/R_\star$ and  limb darkening parameters, as well as the zero point, slope in time, and jitter in each channel. The posteriors are sampled with the Hamiltonian Monte Carlo No-U-Turn Sampler \citep{HoffmanGelman:14} implemented in \texttt{NumPyro} \citep{numpyro} using four chains with $1{,}500$ tuning steps and $1{,}500$ retained draws per chain. 

We bin both datasets to match the variable resolution of NIRSpec/PRISM (corresponding to a resolution of $\sim100$ for NIRISS and $\sim300$--$500$ for NIRSpec)  and fit the spectroscopic light curves directly at that resolution. This retains the broad molecular information while providing a consistent basis for future comparisons among instruments and targets.

\section{Atmospheric Retrievals}\label{sec:retrievals}

\subsection{Atmospheric models}

We fit the limb-averaged spectrum with two models, a 1-TP model with inhomogeneous aerosol treatment, and a 2-TP model to model distinctly different planet terminators.
The 1-TP model combines a clear and cloudy component that share a single TP profile, with their transit depths linearly combined with a relative contribution set by a fitted cloud-covering fraction \citep{Line:16}. The 2-TP model instead gives each component its own TP profile and aerosol properties and combines their transit depths with a fitted weight \citep{MacDonald:20, Welbanks:22}. Both approaches retain a shared metallicity and carbon-to-oxygen (C/O) ratio between component. In this study, we opt for modified chemical equilibrium retrievals (as described below), since free-chemistry retrievals can result in biased abundances and thus biased metallicity constraints due to non-detections of potentially dominant species (e.g., CO), resulting in prior-dominated posteriors that can be orders of magnitude lower than their expected values \citep{line_systematics_2013, Savel:26, radica_supersolar_2026}. 

At each likelihood evaluation in the retrievals, \texttt{FastChem} \citep{Fastchem} computes equilibrium abundances from the fitted metallicity and C/O ratio using the solar composition of \citet{Asplund:21}. We independently fit a vertically constant SO$_2$ abundance, since it is a non-equilibrium photochemical product, in addition to the equilibrium chemistry parameters. We note that although SO$_2$ is not a prominent opacity source in WASP-94Ab, its inclusion naturally extends our framework for use with colder planets where it not only may be present, but also be detectably different between terminators \citep[e.g., WASP-39b;][]{tsai_photochemically_2023}. 

The full opacity set includes H$_2$O, CO$_2$, CO, CH$_4$, Na, K, H$_2$S, HCN, NH$_3$, and SO$_2$, together with H$_2$--H$_2$/H$_2$--He collision-induced absorption and Rayleigh scattering \citep{Polyansky:18,Yurchenko:20,Li:15,Yurchenko:24,Allard:19,Allard:16,Azzam:16,Barber:14,Coles:19,Underwood:16,Zhang:25}. Cross-sections are sampled at $R=20{,}000$ before integration over the observed bins. The transmission calculation follows spherical slant paths through 100 layers uniformly spaced in $\log P$ from $10^{-9}$ to $10^3$ bar, with the planetary radius defined at 1 bar.

Each atmospheric column adopts the irradiated pressure--temperature profile of \citet{Guillot:10}, for which we fit the limb temperature ($T_{\rm irr}$), visible-to-infrared opacity ratio ($\gamma$), internal temperature ($T_{\rm int}$) and thermal opacity ($\kappa_{\rm th}$).
Clouds and hazes enter through a cloud-top pressure ($P_{\rm cloud}$), scattering amplitude ($A_{\rm scatt}$) and scattering slope ($s$). We additionally fit the planet mass ($M_{\rm p}$), 1-bar radius ($R_{\rm p}$), metallicity, C/O and detector offsets.

The 1-TP model adds a cloud covering fraction ($f_{\rm cloud}$), linearly combining clear and cloudy spectra computed from a single thermal profile \citep{Line:16}. The 2-TP model instead fits $T_{\rm irr}$, $\gamma$, the cloud and haze parameters, and an SO$_2$ abundance for each limb, while sharing $T_{\rm int}$ and $\kappa_{\rm th}$; the two spectra are combined with a cold contribution fraction ($f_{\rm cold} = 1 - f_{\rm hot}$). Sharing the deep-atmosphere parameters, which transmission spectroscopy scarcely probes, provides a physically motivated anchor, while the separate irradiation parameters let the columns diverge across the radiative pressures sampled by the data; their common elemental composition then yields different local equilibrium abundance profiles. We emphasize that these weights describe only the spectral mixture, and cannot constrain the latitudes and longitudes from which each component originates. We provide the priors in Table~\ref{tab:priors}.

\begin{table}[t]
\centering
\caption{Retrieval Priors}
\label{tab:priors}
\footnotesize
\begin{tabular}{lll}
\toprule
Parameter & Prior & Model \\
\midrule
$M_{\rm p}/M_{\rm J}$ 
    & $\mathcal{N}(0.456,0.035)$ & Both \\
$R_{\rm p}$ 
    & $\mathcal{U}(0.85,1.15)\times1.58R_{\rm J}$ & Both \\
$\log_{10}(\zsun)$ 
    & $\mathcal{U}(-1,3)$ & Both \\
C/O 
    & $\mathcal{U}(0.1,2.0)$ & Both \\
Detector offset (ppm)
    & $\mathcal{U}(-1000, 1000)$ & Both \\
$\log_{10}X_{\rm SO_2}$ 
    & $\mathcal{U}(-12,-1)$ & Both \\
$T_{\rm int}$ (K) 
    & $\mathcal{U}(30,630)$ & Both \\
$\log_{10}[\kappa_{\rm th}/(\mathrm{cm^2\,g^{-1}})]$ 
    & $\mathcal{U}(-2,0.5)$ & Both \\
$T_{\rm irr}$ (K) 
    & $\mathcal{U}(300,3000)$ & Both \\
$\log_{10}\gamma$ 
    & $\mathcal{U}(-2,0.5)$ & Both \\
$\log_{10}(P_{\rm cloud}/\mathrm{bar})$ 
    & $\mathcal{U}(-6,3)$ & Both \\
$\log_{10}A_{\rm scatt}$ 
    & $\mathcal{U}(-4,8)$ & Both \\
$s$ 
    & $\mathcal{U}(-2,15)$ & Both \\
$f_{\rm cloud}$ 
    & $\mathcal{U}(0,1)$ & 1-TP \\
$f_{\rm cold}$ 
    & $\mathcal{U}(0,1)$ & 2-TP \\
\bottomrule
\end{tabular}
\par\smallskip
\parbox{0.98\columnwidth}{\footnotesize
\textit{Note.} $\mathcal{N}$ and $\mathcal{U}$ denote normal and uniform priors.}
\end{table}

We define SOSS as the reference transit-depth level and fit two additive offsets that place NRS1 and NRS2 relative to it, each with a uniform $\pm1000$ ppm prior. Because the NIRISS and NIRSpec visits were obtained four months apart, their joint interpretation assumes that the terminator's spectral morphology remained stable between epochs. We also test a transit light source (TLS) contamination model \citep{Rackham:18}. These fits provide no evidence for stellar heterogeneity---consistent with the previous analyses---and including TLS contamination does not materially change the planetary inference. We therefore adopt the homogeneous-photosphere retrievals. 

Each atmospheric model is sampled with \texttt{PyMultiNest} \citep{Pymultinest}, using 1000 live points \citep{Skilling:2006,Feroz:09}. We report the preference for 2-TP through the log-evidence difference ($\Delta\ln\mathcal{Z}=\ln\mathcal{Z}_{2-{\rm TP}}-\ln\mathcal{Z}_{1-{\rm TP}}$) and the Bayes factor ($K=\exp(\Delta\ln\mathcal{Z})$). Finally, we performed 1-TP and 2-TP retrievals with a modified version of \texttt{petitRADTRANS} to test whether the metallicity shift depends on the radiative-transfer framework (Appendix~\ref{app:prt}), but find that our conclusions remain unchanged. 

\subsection{Accelerated retrieval implementation}

Both our 1-TP and 2-TP retrievals are implemented in \platon\ 7.0, which replaces the \texttt{NumPy} calculations in \platon\ 6.0 with a \texttt{JAX} implementation of the forward model and likelihood, enabling compilation and GPU execution \citep{PLATON7,JAX}. Moreover, we default to 32-bit floating point precision. Appendix~\ref{app:platonvalidation} presents the numerical implementation and forward model parity tests. The speedups exceed a factor of 100 in every configuration benchmarked for this work, and they rise to several hundred in the fastest cases. The accelerated implementation is available in the public version of \platon\footnote{\url{https://github.com/ideasrule/platon/}}.

This acceleration is especially consequential for the 2-TP retrievals, where nested sampling repeatedly evaluates both atmospheric columns across a large parameter space. The broad-wavelength, 21-parameter retrieval completes in one day on an NVIDIA A100 GPU; applying the measured \platon\ 6.0 evaluation time to the same likelihood count gives an estimated runtime with the \texttt{NumPy} version of 187 days. A consumer-grade RTX 5070 Ti completes the accelerated retrieval approximately 25\% slower. 

\section{Results}\label{sec:results}

\subsection{A 2-TP model is statistically favored over 1-TP}

\begin{figure*}[t!]
\centering
\includegraphics[width=\textwidth]{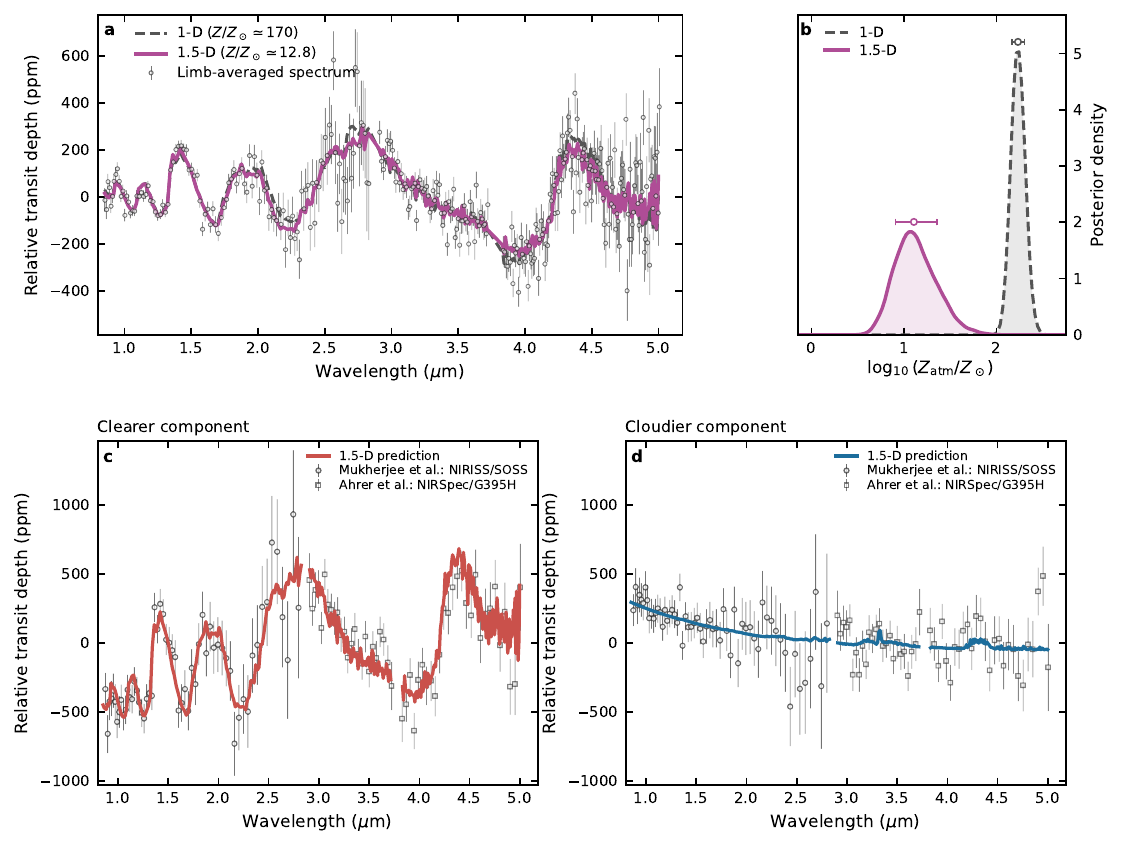}
\caption{The limb-averaged analysis and its comparison with independently published reductions. Panel (a) shows the limb-averaged $0.85$--$5.0\,\mu{\rm m}$ NIRISS/SOSS and NIRSpec/G395H spectrum with the best-fitting 1-TP and 2-TP models. Panel (b) shows the corresponding atmospheric metallicity posteriors relative to the Sun. Panels (c) and (d) compare the clearer and cloudier components predicted by the limb-averaged 2-TP retrieval with the published NIRISS/SOSS limb spectra of \citetalias{Mukherjee_2026a} (circles) and NIRSpec/G395H limb spectra of \citetalias{Ahrer:25a} (squares). We note that the limb-resolved spectra presented in panels c and d are not used in our retrievals, i.e., the red and blue models are not fits to the data shown in these panels, but rather the two components extracted directly from the spectrum plotted in panel a.}
\label{fig:limbaveraged}
\end{figure*}
 
The standard 1-TP retrieval fits the limb-averaged spectrum reasonably well (Figure~\ref{fig:limbaveraged}a), but it returns a metallicity of $\log_{10}(\zsun)=\onedlogz$, or $\onedlinearz\times$ solar, which is difficult to reconcile with the planet's interior structure. A metallicity this high would require an atmosphere more enriched in metals than the bulk interior, a configuration disfavored by interior-structure modeling \citep{thorngren_connecting_2019}. The 2-TP retrieval relieves this tension: it is favored over the 1-TP model by $\Delta\ln\mathcal{Z}\simeq\combineddeltaevidence$, a Bayes factor of $K\simeq\combinedbayesfactor$ ($\combinedbayesfactor:1$ odds) despite its six additional parameters, and it lowers the metallicity by more than an order of magnitude to $\twodlogz$, or $\twodlinearz\times$ solar. We found the results are robust to the use of isothermal profiles instead of Guillot profiles, and when the two 2-TP components have separate thermal opacities, which both return unconstrained values; these tests and results are reported quantitatively in Appendix~\ref{app:D}. We show the joint posterior plots for both the 1-TP and 2-TP retrievals in Appendix Figures~\ref{app:corner1D} and \ref{app:corner1.5D}. The clear component's temperature is well constrained, $T_{\rm hot}=1090_{-70}^{+90}$\,K at 1\,mbar, whereas the aerosol-obscured component's temperature only has an upper limit, which depends on the prior parameterization (Section~\ref{ssec:coldcomponent}). Because the metallicity and C/O are set by the clear component's molecular features, this does not affect the inferred composition. 
 
\begin{figure*}[t!]
\centering
\includegraphics[width=\textwidth]{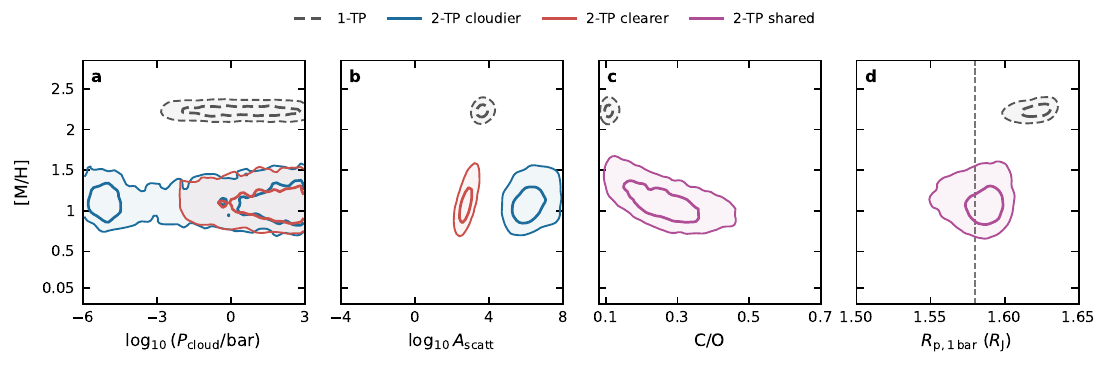}
\caption{Joint posterior constraints from the limb-averaged spectrum. From (a) to (d), the panels show the opaque cloud-top pressure, haze-scattering amplitude, C/O, and 1-bar radius against metallicity. (a) and (b) compare the 1-TP model with the independently retrieved hot and cold components of the 2-TP model. The cold component requires substantially stronger haze scattering, whereas its gray cloud-top pressure remains weakly constrained because the haze already mutes the spectrum; a large $P_{\rm cloud}$ places the gray deck deep in the atmosphere and makes it spectrally irrelevant. C/O, radius, and metallicity are shared between the two components of the 2-TP model. Contours enclose 39.3\% and 86.5\% of the posterior probability. The dashed line in panel (d) represents the white-light planet radius.}
\label{fig:coverage}
\end{figure*}

A better fit and a more physically plausible value do not, on their own, establish that the 2-TP metallicity is correct, so we compare it against an independent benchmark. \citetalias{Mukherjee_2026a} retrieved on the NIRISS/SOSS spectra of the morning and evening limbs jointly and obtained $[\mathrm{M/H}]=0.46\pm0.36$, or $2.9^{+3.7}_{-1.6}\times$ solar. Because that retrieval fits each limb's absolute transit depth, we take it as the more reliable estimate of the planet's true metallicity. Our 2-TP value lies within $\sim1.6\sigma$ of this benchmark, while the 1-TP value differs by $4.8\sigma$ (Figure~\ref{fig:coverage}). The remaining offset between our 2-TP result and \citetalias{Mukherjee_2026a} is likely driven by our inclusion of NIRSpec/G395H alongside NIRISS/SOSS, which adds constraints on the carbon-bearing species \citep{batalha17,Heinke:26}. 

More fundamentally, only the low-metallicity solution is compatible with the limb-resolved data at all because a 1-TP atmosphere produces a single transmission spectrum by construction, and cannot simultaneously reproduce two limbs with different aerosol and thermal structure, whereas the two components of the 2-TP retrieval---obtained without any limb-resolved information---independently reproduce both limbs, as we show next.

\subsection{The 2-TP model components reproduce the limb-resolved spectra}
\label{ssec:limbrecovery}

The 2-TP retrieval of the limb-averaged spectrum identifies a clearer component and a aerosol-muted component, with a scattering amplitude $\sim$$\limbdeltascatt$ dex larger than that of the warmer component. The two contribute in nearly equal measure, with $f_{\rm cold}=\limbweight$, consistent with one component arising from each terminator hemisphere. 

We next compare the two spectral components of our 2-TP model to the published limb-resolved measurements from \citetalias{Mukherjee_2026a} and \citetalias{Ahrer:25a}. For the comparison, we normalize each limb spectrum by removing the inverse-variance-weighted mean (separately for NIRISS and NIRSpec) thereby preserving the spectral shape while removing any offset in the absolute transit depth. The clearer component reproduces the feature-rich evening limb across the full $0.85$--$5.0\,\mu{\rm m}$ range, including the H$_2$O bands measured with NIRISS and the CO$_2$ and candidate CO absorption measured with G395H ($\chi^2_\nu=\eveninglimbredchisq$; Figure~\ref{fig:limbaveraged}(c)). The cloudier component similarly reproduces the aerosol-muted morning limb across both instruments ($\chi^2_\nu=\morninglimbredchisq$; Figure~\ref{fig:limbaveraged}(d)). Combined across both limbs and instruments, the predicted components fit the full limb-resolved dataset with $\chi^2_\nu=\jointlimbredchisq$. 

\subsection{Constraints on the cloudier component's properties}
\label{ssec:coldcomponent}

Because the cloudier component is nearly featureless, it is worth asking whether our model connects it to a temperature at all. Its continuum and slope are produced by the parameteric haze, whose cross section is $A_{\rm scatt}$ times the Rayleigh cross section at $\lambda_0=1\,\mu$m, scaled as ($\lambda/\lambda_0)^{-s}$. This also means the continuum and slope are not produced by collision-induced absorption (CIA) or the Rayleigh scattering from the gas. The haze cross section exceeds the gas Rayleigh cross section by $\sim10^6$ and places the cloudier component's slant photosphere near 1\,$\mu$bar, where CIA, which scales with the square of the density, is negligible; removing CIA leaves the spectrum unchanged. For the haze and gas opacities, an isothermal column has \citep{lecavelier_des_etangs_rayleigh_2008},
\begin{align}
sH\simeq \frac{d R(\lambda)}{d\ln\lambda},
\end{align}
where $R(\lambda)$ is the transit radius, $H=k_BT/(\mu m_H g)$ and $\mu\approx2.6$ is fixed by the shared composition. The slope therefore constrains only $sH$ \citep{Heng:16}, and $s$ is a fitted parameter because condensates do not need to scatter with the Rayleigh slope \citep{Kitzmann:18}. The continuum level trades against $A_{\rm scatt}$ and the shared 1-bar radius $R_p$, and a larger $A_{\rm scatt}$ mutes any residual bands. The posterior show these degeneracies (Figure~\ref{appfig:E1}), where $T_{\rm cold}$ rises from $\lesssim400$ to $\gtrsim900$\,K at 1\,mbar, $s_{\rm cold}$ falls roughly as $1/T$, $A_{\rm scatt,cold}$ rises by an order of magnitude, and $R_{\rm{p}}$ shrinks. 

The posterior disfavors warmer solutions once these degeneracies are driven to the edge of their priors. A warmer column has a larger scale height and therefore larger residual bands, so it needs a larger haze amplitude to stay featureless, and the shared $R_p$ must shrink to hold the continuum level. Once the prior bounds are met, we obtain a 2$\sigma$ upper limit of $T_{\rm cold}<670$\,K at 1\,mbar (Figure~\ref{appfig:E1}a). With $T_{\rm hot}$ set by the molecular bands, the corresponding $2\sigma$ lower limit on the limb temperature contrast is $\Delta T>390$\,K. There is no lower limit on $T_{\rm cold}$, and the posterior median ($\approx 450$\,K) reflects the 300\,K lower bound of the $T_{\rm irr}$ prior. These limits are conditional on the haze parameterization and its priors. The limits on $T_{\rm cold}$ and $\Delta T$ are therefore covariant with the aerosol assumptions. We leave a treatment with physically motivated aeosol models to future work. The composition is unaffected as the metallicity and C/O are set by the clearer component's bands and are uncorrelated with $T_{\rm cold}$ (Spearman $\rho=-0.08$ and $-0.07$).

\section{Discussion}\label{sec:discussion}

\subsection{Why one atmospheric column biases the composition}

\begin{figure*}[t!]
\centering
\includegraphics[scale=0.8, keepaspectratio]{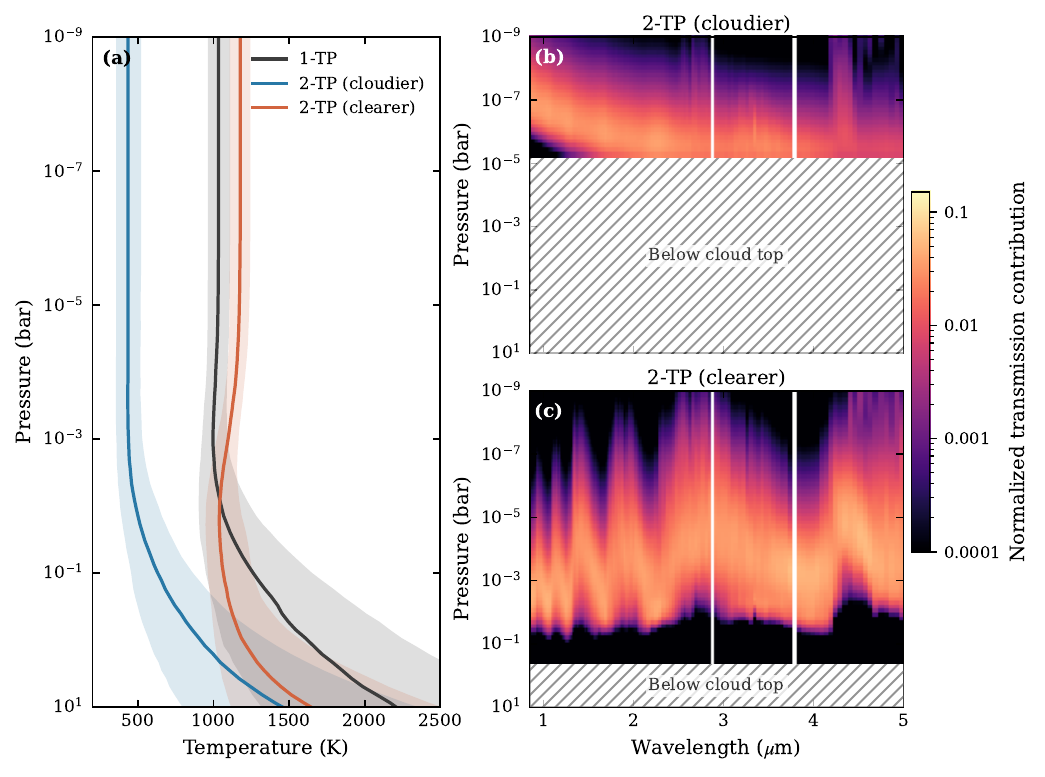}
\caption{Thermal structures and pressure regions probed by the retrieved models. Panel (a) compares the posterior median pressure--temperature profiles of the 1-TP model and the cloudier and clearer components of the 2-TP model; shaded regions enclose the 68\% credible intervals. Panels (b) and (c) show transmission contribution functions evaluated at the best-fitting 2-TP solution. The upper boundary of each hatched region marks the component's cloud top. White vertical strips mark gaps between the observed wavelength ranges.}
\label{fig:tpcontribution}
\end{figure*}

Our retrieval results identify the driving factors behind the metallicity discrepancy between 1-TP and \mbox{2-TP} interpretations. Clouds transported through a hot Jupiter's circulation can preferentially form or persist along the cooler limb, raise its effective continuum, and evaporate as material reaches warmer longitudes \citep{Parmentier:16,Roman:19, Lines:19,Powell:19,roman_clouds_2021, Mukherjee_2026a}. In WASP-94Ab, the aerosol-muted component supplies the elevated, weakly structured continuum while the clearer component contributes most of the molecular modulation. A 1-TP atmosphere must reproduce both signatures with one temperature-pressure profile and one cloud prescription, driving the solution toward a high mean molecular weight that reduces the scale height together with compensating shifts in radius and aerosol properties. This general pattern appears in the joint posteriors of both retrievals (Figure~\ref{fig:coverage}) %; Appendix Figures~\ref{app:corner1D} and \ref{app:corner1.5D}).
Once the two chords are allowed with independent thermal and aerosol structures, the continuum and molecular bands can instead arise from different atmospheric regions---as shown by the pressure ranges probed by each component in Figure~\ref{fig:tpcontribution}---and which provides an improved fit overall.

Critically, WASP-94Ab is one of the few planets for which morning--evening asymmetry is independently confirmed via limb-resolved observations \citep{Mukherjee_2026a, Ahrer:25a}. Because that asymmetry is already established as a ground truth, we can assert that the 2-TP retrieval's preference for two atmospheric columns reflects a genuine physical feature of the planet, rather than a statistical artifact of the additional retrieval freedom. 

\subsection{Limb-averaged spectra as predictors of limb asymmetry}

The limb-averaged retrieval constrains the shared composition and relative contributions of (any) two strongly contrasting atmospheric components.  For tidally locked hot Jupiters, this is consistent with predictions that morning--evening asymmetries remain encoded in limb-averaged spectra \citep{MacDonald:20,Welbanks:22}.  However, recovering two distinct spectral components does not necessarily mean that they arise from morning and evening limbs (as opposed to, e.g., pole and equator). Furthermore, exchanging the locations of these components around the terminator leaves the limb-averaged spectrum unchanged. 

It is only through ingress- and egress-resolved analyses that the geometric information is added by breaking the degeneracy between the morning and evening limbs and measuring the absolute, wavelength-dependent radius contrast between them \citep{EspinozaJones:21,JonesEspinoza:22,GrantWakeford:23}. Together, the two approaches (time resolved spectroscopy vs.\ the 2-TP retrievals developed in this work) point toward a practical observing strategy. Broad, limb-averaged spectra can first constrain composition and flag systems with statistical evidence for terminator asymmetry. Targeted ingress- and egress-resolved analyses on high S/N data sets can then be used to assign the inferred components to individual limbs and test the predicted spectra directly.

While this \textit{Letter} was under review, \citet{Mukherjee_2026} published a complementary analysis of WASP-94Ab that incorporates the transit time-shift spectrum alongside the transmission spectrum. Consistent with our results, they find preference for two distinct limbs---one cloudier and one clearer. They infer a metallicity 0.8\,dex lower than our 2-TP value, likely reflecting differences in retrieval setup (specifically, free chemistry versus our chemical-equilibrium framework; Section~\ref{sec:retrievals}).

\subsection{Implications for the transiting exoplanet population}

Most transmission spectroscopy studies fit a single, limb-averaged atmospheric column---the 1-TP retrieval. If morning--evening asymmetry is common among close-in giant planets, as circulation models predict, then the metallicity, temperature, and cloud properties inferred from such fits could be biased for essentially any transiting planet. Here we show that our 2-TP retrievals can test for, and alleviate this bias in the benchmark case of WASP-94Ab. If this behavior generalizes, we argue for routinely testing 1-TP against 2-TP retrievals wherever accurate compositions are needed. 

\begin{figure}[b!]
\centering
\includegraphics[width=\columnwidth]{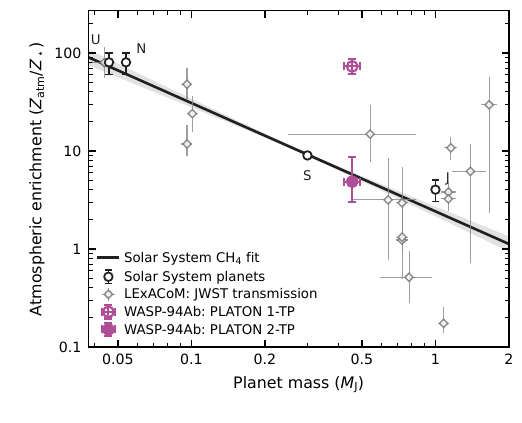}
\caption{Mass--metallicity context for WASP-94Ab and the current JWST transmission sample. Open circles show the Solar System CH$_4$-based C/H metallicities compiled by \citet{Atreya:22}, with the black curve showing the fitted relation. Gray diamonds show individual JWST metallicity constraints from the LExACoM database \citep{Lothringer:26}, normalized by the corresponding host-star metallicity. Multiple entries for the same planet represent separate literature analyses.
The empty and filled circle show the fiducial \platon\ 1-TP and 2-TP interpretations of WASP-94Ab.}
\label{fig:massmetal}
\end{figure}

The factor-of-$\metallicityratio$ abundance shift between 1-TP and 2-TP analyses is already large enough to change the formation context assigned to WASP-94Ab. Using $[\mathrm{Fe/H}]_\star=0.37\pm0.01$ \citep{Sousa:21} as a proxy for the host enrichment gives $\twodhostz\times$ stellar in 2-TP and $\onedhostz\times$ stellar in 1-TP. A log-linear fit to the Solar System giant-planet C/H enrichments compiled by \citet{Atreya:22} predicts approximately $\wasptrendz\times$ stellar for WASP-94Ab ($M_{\rm p}=0.456\pm0.035\,M_{\rm J}$; Figure~\ref{fig:massmetal}). The 2-TP result lies on that expectation, whereas the 1-TP value is approximately $\onedtrendratio$ times higher, in clear tension with the solar system trend. 

The LExACoM compilation of JWST transmission spectra analyzed with 1-TP techniques \citep{Lothringer:26} also contains several metallicity measurements that lie far from the Solar System relation. Our work demonstrates that at least some of the apparently metal-rich outliers may result partly from fitting an asymmetric terminator with a single atmospheric column. This does not imply that multidimensional retrievals will move every planet toward the solar system mass-metallicity relation, or even bias every atmosphere in the same direction. If terminator asymmetry varies systematically with temperature, gravity, or cloud regime, however, retrieval dimensionality could affect both the slope and scatter inferred for atmospheric mass--metallicity relations \citep{MacDonald:20,Welbanks:22}. Comparative \mbox{1-TP} and 2-TP reanalyses of the broad-wavelength JWST sample can test that possibility.

\section{Conclusions}\label{sec:conclusions}

Our analysis of the $0.85$--$5.0\,\mu{\rm m}$ limb-averaged spectrum of WASP-94Ab leads to three main conclusions:

\begin{itemize}
\item The limb-averaged spectrum favors the 2-TP retrieval over the 1-TP retrieval with a Bayes Factor of $K=\combinedbayesfactor$, and lowers the inferred metallicity by over an order of magnitude from $\onedlinearz$ to $\twodlinearz \times$ solar.  The 2-TP value resolves inconsistencies with bulk metallicity constraints from interior structure models \citep[e.g.,][]{thorngren_connecting_2019} and also broadly agrees with the NIRISS-only limb-resolved result of \citetalias{Mukherjee_2026a}.

\item Without using any limb-resolved data products, the 2-TP retrieval separates a clearer component from a cloudier component and reproduces the published evening- and morning-limb spectra across both NIRISS and NIRSpec \citep{Mukherjee_2026a,Ahrer:25a}, with $\chi^2_\nu=\eveninglimbredchisq$ and $\morninglimbredchisq$, respectively. The retrieval also constrains the cloudier component's effective contribution to $f_{\rm cold}=\limbweight$, indicating that each component is consistent with representing a single hemisphere. The clearer component's temperature is $1090_{-70}^{+90}$\,K at 1\,mbar, but the aerosol-muted component is only bounded as an upper limit ($<670$\,K at $2\sigma$, so $\Delta T>390$\,K), which is conditional on the haze parameterizations and its prior ranges. The strong agreement between 2-TP and limb-resolved analyses demonstrates that the spectral signatures of terminator asymmetry can be recovered from the limb-averaged spectrum.

\item Our results demonstrate that it is possible to recover debiased metallicities that account for distinct two-limb spectra without requiring a phase-resolved analysis. This opens the door to metallicity studies of a much broader range of planets than the subset amenable to limb-resolved spectroscopy, ultimately enabling population-level studies.

\end{itemize}

These calculations are made practical by \texttt{PLATON 7.0}, whose \texttt{JAX}-accelerated implementation reduces the fiducial 2-TP retrieval from an estimated 187 days to one day (on an NVIDIA A100 GPU). For WASP-94Ab, both the cloudy and clear components, as well as the lower metallicity first revealed by limb-resolved spectroscopy were present in the limb-averaged spectrum all along.

\section*{Data and Code Availability}

The JWST observations are publicly available from the Mikulski Archive for Space Telescopes under programs GO~3154 and GO~5924. The data reduction pipeline can be found at \url{https://github.com/radicamc/exoTEDRF}. The accelerated 2-TP implementation of \platon\ is available at \url{https://github.com/ideasrule/platon/}.  The light-curve models use \texttt{jaxoplanet}, available at \url{https://github.com/exoplanet-dev/jaxoplanet}. Any further code is available upon reasonable request to the corresponding author. The reduced spectra, posterior samples used in this Letter and retrieval setup is available in a Zenodo repository \url{https://zenodo.org/records/23247451}.

\begin{acknowledgments}
This analysis was based on observations made with the NASA/ESA/CSA James Webb Space Telescope. The data were obtained from the Mikulski Archive for Space Telescopes at the Space Telescope Science Institute, which is operated by the Association of Universities for Research in Astronomy, Inc., under NASA contract NAS~5-03127. These observations are associated with programs GO~3154 and GO~5924.

M.R. acknowledges support from a Natural Sciences and Engineering Research Council of Canada Postdoctoral Fellowship. T.R.F and E.M.-R.K. acknowledge support from JWST program 5275, provided by NASA through a grant from the Space Telescope Science Institute.

T.R.F thanks Evert Nasedkin for helpful conversations regarding the usage of petitRADTRANS. 

The data presented in this article were obtained from the Mikulski Archive for Space Telescopes (MAST) at the Space Telescope Science Institute. The specific observations analyzed can be accessed via \dataset[doi:10.17909/1yms-w890]{https://doi.org/10.17909/1yms-w890}.

This work used Delta at the University of Illinois Urbana-Champaign through allocation PHY260183 from the Advanced Cyberinfrastructure Coordination Ecosystem: Services \& Support (ACCESS) program, which is supported by National Science Foundation grants 2138259, 2138286, 2138307, 2137603, and 2138296 \citep{ACCESS}.

This work was completed in part with resources provided by the University of Chicago’s Research Computing Center. 

We acknowledge the use of A.I. tools (ChatGPT, Claude) for assistance in coding and refinement of ideas. The authors take all responsibility for the content presented in the paper and all text was human-written. 

\end{acknowledgments}

\begin{contribution}

T.R.F. led the data analysis, including the reduction stages, the light curve fitting, the development of the 2-TP retrieval configuration in \texttt{PLATON 7.0}, interpretation of results, and manuscript development.
M.R. assisted T.R.F. during the initial reductions of the data, including the use of the optimizer, provided guidance in light curve fitting, and contributed significantly to both the interpretation of the results and the subsequent manuscript.
M.Z. led the development of \texttt{PLATON 7.0}, which is used as the backbone of the research conducted.
E.M.-R.K. \& J.L.B. supervised T.R.F. during the data analysis stages, assisted in the development of the 2-TP configuration and interpretation of results, and contributed significantly to the writing of the manuscript.
E.-M.A. modeled the independent limb-resolved spectra and provided feedback on the interpretation of the results.
A.S. assisted in the retrieval configurations and contributed to the interpretation of asymmetries.
P.S.D. led the initial development of the reduction optimizer framework.
All authors read and approved the final manuscript.
\end{contribution}

\software{
\texttt{astropy} \citep{astropy:2013, astropy:2018}, 
\texttt{exoTEDRF} \citep{radica_awesome_2023, feinstein_early_2023, Radica:24}, 
\texttt{PLATON} \citep{Zhang:20,Zhang:25, Fairnington:inprep}, \texttt{FastChem} \citep{Fastchem}, 
\texttt{jaxoplanet} \citep{jaxoplanet}, 
\texttt{jwst} \citep{bushouse_2023},
\texttt{matplotlib} \citep{Hunter:2007},
\texttt{numpy} \citep{harris2020array},
\texttt{NumPyro} \citep{numpyro}, 
\texttt{PyMultiNest} \citep{Pymultinest},
\texttt{scipy} \citep{2020SciPy-NMeth},
\texttt{Claude.ai Fable 5} \citep{claude5fable},
\texttt{ChatGPT 5.6 Sol} \citep{chatgpt56sol}
}

\bibliographystyle{aasjournalv7}
\bibliography{references}

@ARTICLE{Nixon:22,
       author = {{Nixon}, Matthew C. and {Madhusudhan}, Nikku},
        title = "{Aura-3D: A Three-dimensional Atmospheric Retrieval Framework for Exoplanet Transmission Spectra}",
      journal = {\apj},
         year = 2022,
        month = aug,
       volume = {935},
       number = {2},
          eid = {73},
        pages = {73},
          doi = {10.3847/1538-4357/ac7c09},
archivePrefix = {arXiv},
       eprint = {2201.03532},
 primaryClass = {astro-ph.EP},
       adsurl = {https://ui.adsabs.harvard.edu/abs/2022ApJ...935...73N}
}

@article{Powell:19,
  author =        {{Powell}, Diana and {Louden}, Tom and
                   {Kreidberg}, Laura and {Zhang}, Xi and {Gao}, Peter and
                   {Parmentier}, Vivien},
  journal =       {\apj},
  pages =         {170},
  title =         {{Transit Signatures of Inhomogeneous Clouds on Hot
                   Jupiters}},
  volume =        {887},
  year =          {2019},
  doi =           {10.3847/1538-4357/ab55d9},
  eid =           {170},
}

@ARTICLE{pRT3,
       author = {{Blain}, Doriann and {Molli{\`e}re}, Paul and {Nasedkin}, Evert},
        title = "{SpectralModel: a high-resolution framework for petitRADTRANS 3}",
      journal = {The Journal of Open Source Software},
         year = 2024,
        month = oct,
       volume = {9},
       number = {102},
          eid = {7028},
        pages = {7028},
          doi = {10.21105/joss.07028},
       adsurl = {https://ui.adsabs.harvard.edu/abs/2024JOSS....9.7028B}
}

@unpublished{Fairnington:inprep,
  author       = {Fairnington, Tyler R. and others},
  title        = {manuscript on optimizer},
  howpublished = {in preparation},
  year         = {2026}
}

@article{Showman:09,
doi = {10.1088/0004-637X/699/1/564},
url = {https://doi.org/10.1088/0004-637X/699/1/564},
year = {2009},
month = {jun},
publisher = {The American Astronomical Society},
volume = {699},
number = {1},
pages = {564},
author = {Showman, Adam P. and Fortney, Jonathan J. and Lian, Yuan and Marley, Mark S. and Freedman, Richard S. and Knutson, Heather A. and Charbonneau, David},
title = {ATMOSPHERIC CIRCULATION OF HOT JUPITERS: COUPLED RADIATIVE-DYNAMICAL GENERAL CIRCULATION MODEL SIMULATIONS OF HD 189733b and HD 209458b},
journal = {The Astrophysical Journal}
}

@ARTICLE{Rauscher:12,
       author = {{Rauscher}, Emily and {Menou}, Kristen},
        title = "{A General Circulation Model for Gaseous Exoplanets with Double-gray Radiative Transfer}",
      journal = {\apj},
         year = 2012,
        month = may,
       volume = {750},
       number = {2},
          eid = {96},
        pages = {96},
          doi = {10.1088/0004-637X/750/2/96},
archivePrefix = {arXiv},
       eprint = {1112.1658},
 primaryClass = {astro-ph.EP},
       adsurl = {https://ui.adsabs.harvard.edu/abs/2012ApJ...750...96R}
}

@article{pRT4, doi = {10.21105/joss.05875}, url = {https://doi.org/10.21105/joss.05875}, year = {2024}, publisher = {The Open Journal}, volume = {9}, number = {96}, pages = {5875}, author = {Nasedkin, Evert and Mollière, Paul and Blain, Doriann}, title = {Atmospheric Retrievals with petitRADTRANS}, journal = {Journal of Open Source Software} }

@ARTICLE{Heinke:26,
       author = {{Heinke}, L. and {Min}, M. and {Bouwman}, J. and {Crouzet}, N. and {Konings}, T. and {Decin}, L. and {Waters}, L.~B.~F.~M. and {Lagage}, P.-O. and {Henning}, T. and {Palmer}, P.~I. and {Edwards}, B. and {Pye}, J.~P. and {G{\"u}del}, M. and {Absil}, O. and {Barrado}, D. and {Cossou}, C. and {Glasse}, A. and {Glauser}, A.~M. and {{\"O}stlin}, G. and {Whiteford}, N. and {Ray}, T.~P.},
        title = "{Information content of JWST transmission spectroscopy of the exoplanet HAT-P-12b from the optical to the mid-infrared}",
      journal = {\aap},
         year = 2026,
        month = jun,
       volume = {710},
          eid = {A305},
        pages = {A305},
          doi = {10.1051/0004-6361/202557839},
archivePrefix = {arXiv},
       eprint = {2604.01219},
 primaryClass = {astro-ph.EP},
       adsurl = {https://ui.adsabs.harvard.edu/abs/2026A&A...710A.305H}
}

@article{Caldas:19,
  author =        {{Caldas}, Anthony and {Leconte}, J{\'e}r{\^o}me and
                   {Selsis}, Franck and others},
  journal =       {\aap},
  pages =         {A161},
  title =         {{Effects of a Fully 3D Atmospheric Structure on
                   Exoplanet Transmission Spectra: Retrieval Biases due
                   to Day--Night Temperature Gradients}},
  volume =        {623},
  year =          {2019},
  doi =           {10.1051/0004-6361/201834384},
  eid =           {A161},
}

@article{Pluriel:22,
  author =        {{Pluriel}, William and {Zingales}, Tiziano and
                   {Leconte}, J{\'e}r{\^o}me and {Parmentier}, Vivien},
  journal =       {\aap},
  pages =         {A42},
  title =         {{Toward a Multidimensional Analysis of Transmission
                   Spectroscopy. II. Day--Night-induced Biases in
                   Retrievals from Hot to Ultrahot Jupiters}},
  volume =        {658},
  year =          {2022},
  doi =           {10.1051/0004-6361/202141943},
  eid =           {A42},
}

@article{Line:16,
  author =        {Line, Michael R. and Parmentier, Vivien},
  journal =       {The Astrophysical Journal},
  month =         {mar},
  number =        {1},
  pages =         {78},
  publisher =     {The American Astronomical Society},
  title =         {THE INFLUENCE OF NONUNIFORM CLOUD COVER ON TRANSIT
                   TRANSMISSION SPECTRA},
  volume =        {820},
  year =          {2016},
  doi =           {10.3847/0004-637X/820/1/78},
  url =           {https://doi.org/10.3847/0004-637X/820/1/78},
}

@article{MacDonald:20,
  author =        {{MacDonald}, Ryan J. and {Goyal}, Jayesh M. and
                   {Lewis}, Nikole K.},
  journal =       {\apjl},
  month =         apr,
  pages =         {L43},
  title =         {{Why Is It So Cold in Here? Explaining the Cold
                   Temperatures Retrieved from Transmission Spectra of
                   Exoplanet Atmospheres}},
  volume =        {893},
  year =          {2020},
  doi =           {10.3847/2041-8213/ab8238},
  eid =           {L43},
}

@article{Welbanks:22,
  author =        {{Welbanks}, Luis and {Madhusudhan}, Nikku},
  journal =       {\apj},
  month =         jul,
  number =        {1},
  pages =         {79},
  title =         {{On Atmospheric Retrievals of Exoplanets with
                   Inhomogeneous Terminators}},
  volume =        {933},
  year =          {2022},
  doi =           {10.3847/1538-4357/ac6df1},
  eid =           {79},
}

@article{EspinozaJones:21,
  author =        {{Espinoza}, N{\'e}stor and {Jones}, Kathryn},
  journal =       {\aj},
  pages =         {165},
  title =         {{Constraining Mornings and Evenings on Distant
                   Worlds}},
  volume =        {162},
  year =          {2021},
  doi =           {10.3847/1538-3881/ac134d},
  eid =           {165},
}

@article{JonesEspinoza:22,
  author =        {{Jones}, Kathryn and {Espinoza}, N{\'e}stor},
  journal =       {The Journal of Open Source Software},
  pages =         {2382},
  title =         {{catwoman: A Transit Modelling Python Package for
                   Asymmetric Light Curves}},
  volume =        {7},
  year =          {2022},
  doi =           {10.21105/joss.02382},
  eid =           {2382},
}

@article{GrantWakeford:23,
  author =        {{Grant}, David and {Wakeford}, Hannah R.},
  journal =       {\mnras},
  pages =         {5114--5127},
  title =         {{Transmission Strings: A Technique for Spatially
                   Mapping Exoplanet Atmospheres around Their
                   Terminators}},
  volume =        {519},
  year =          {2023},
  doi =           {10.1093/mnras/stac3632},
}

@ARTICLE{Mukherjee_2026a,
       author = {{Mukherjee}, Sagnick and {Sing}, David K. and {Fu}, Guangwei and {Stevenson}, Kevin B. and {Schmidt}, Stephen P. and {Baskett}, Harry and {Mak}, Mei Ting and {McCreery}, Patrick and {Allen}, Natalie H. and {Bennett}, Katherine A. and et al.},
        title = "{Cloudy mornings and clear evenings on a gas giant exoplanet}",
      journal = {Science},
         year = 2026,
        month = may,
       volume = {392},
       number = {6800},
        pages = {858-862},
          doi = {10.1126/science.adx5903},
archivePrefix = {arXiv},
       eprint = {2505.10910},
 primaryClass = {astro-ph.EP},
       adsurl = {https://scixplorer.org/abs/2026Sci...392..858M}
}

@article{Lothringer:26,
  author =        {{Lothringer}, Joshua D. and {Lowson}, Nataliea and
                   {Fu}, Guangwei},
  journal =       {\aj},
  month =         jan,
  number =        {1},
  pages =         {31},
  title =         {{The Library of Exoplanet Atmospheric Composition
                   Measurements: Population-level Trends in Exoplanet
                   Composition with ExoComp}},
  volume =        {171},
  year =          {2026},
  doi =           {10.3847/1538-3881/ae1b8e},
  eid =           {31},
  eprint =        {2510.26785},
}

@article{Chen:25,
  author =        {Chen, Zixin and Ji, Jianghui and Chen, Guo and
                   Yan, Fei and Tan, Xianyu},
  journal =       {The Astronomical Journal},
  month =         {may},
  number =        {6},
  pages =         {294},
  publisher =     {The American Astronomical Society},
  title =         {Asymmetry and Dynamical Constraints in Two-limbs
                   Retrieval of WASP-39 b Inferring from JWST Data},
  volume =        {169},
  year =          {2025},
  doi =           {10.3847/1538-3881/adc803},
  url =           {https://doi.org/10.3847/1538-3881/adc803},
}

@article{radica_supersolar_2026,
	title = {Supersolar {Metallicity} and {Tentative} {Evidence} for {Photochemistry} on {WASP}-96 b from {JWST} and {Ground}-based {VLT} {Transmission} {Spectroscopy}},
	volume = {171},
	issn = {0004-6256, 1538-3881},
	url = {https://iopscience.iop.org/article/10.3847/1538-3881/ae5b9f},
	doi = {10.3847/1538-3881/ae5b9f},
	language = {en},
	number = {5},
	urldate = {2026-05-12},
	journal = {The Astronomical Journal},
	author = {Radica, Michael and Taylor, Jake and Rotman, Yoav and Blecic, Jasmina and Welbanks, Luis and Ahrer, Eva-Maria and Christie, Duncan and Coulombe, Louis-Philippe and Lowry, Gillis and Murphy, Matthew M. and Feinstein, Adina D. and Lafrenière, David and MacDonald, Ryan J. and Mayne, Nathan J. and Tsai, Shang-Min and Zamyatina, Maria},
	month = may,
	year = {2026},
	pages = {314},
}

@article{fu_overcast_2025,
	title = {Overcast {Mornings} and {Clear} {Evenings} in {Hot} {Jupiter} {Exoplanet} {Atmospheres}},
	volume = {989},
	issn = {2041-8205, 2041-8213},
	url = {https://iopscience.iop.org/article/10.3847/2041-8213/adf20f},
	doi = {10.3847/2041-8213/adf20f},
	language = {en},
	number = {1},
	urldate = {2025-12-17},
	journal = {The Astrophysical Journal Letters},
	author = {Fu, Guangwei and Mukherjee, Sagnick and Stevenson, Kevin B. and Sing, David K. and Ashtari, Reza and Mayne, Nathan and Lothringer, Joshua D. and Zamyatina, Maria and Schmidt, Stephen P. and Gascón, Carlos and Allen, Natalie H. and Bennett, Katherine A. and López-Morales, Mercedes},
	month = aug,
	year = {2025},
	pages = {L17},
}

@article{murphy_panchromatic_2025,
	title = {A {Panchromatic} {Characterization} of the {Evening} and {Morning} {Atmosphere} of {WASP}-107 b: {Composition} and {Cloud} {Variations}, and {Insight} into the {Effect} of {Stellar} {Contamination}},
	volume = {170},
	issn = {0004-6256, 1538-3881},
	shorttitle = {A {Panchromatic} {Characterization} of the {Evening} and {Morning} {Atmosphere} of {WASP}-107 b},
	url = {https://iopscience.iop.org/article/10.3847/1538-3881/addf38},
	doi = {10.3847/1538-3881/addf38},
	language = {en},
	number = {1},
	urldate = {2025-09-16},
	journal = {The Astronomical Journal},
	author = {Murphy, Matthew M. and Beatty, Thomas G. and Schlawin, Everett and Bell, Taylor J. and Radica, Michael and Kennedy, Thomas D. and Mehta, Nishil and Welbanks, Luis and Line, Michael R. and Parmentier, Vivien and Greene, Thomas P. and Mukherjee, Sagnick and Fortney, Jonathan J. and Ohno, Kazumasa and Wiser, Lindsey and Arnold, Kenneth and Rauscher, Emily and Edelman, Isaac R. and Rieke, Marcia J.},
	month = jul,
	year = {2025},
	pages = {61},
}

@article{ahrer_tracing_2025,
	title = {Tracing the formation and migration history: molecular signatures in the atmosphere of misaligned hot {Jupiter} {WASP}-94 {A} b using \textit{{JWST}} {NIRSpec}/{G395H}},
	volume = {540},
	copyright = {https://creativecommons.org/licenses/by/4.0/},
	issn = {0035-8711, 1365-2966},
	shorttitle = {Tracing the formation and migration history},
	url = {https://academic.oup.com/mnras/article/540/3/2535/8137879},
	doi = {10.1093/mnras/staf819},
	language = {en},
	number = {3},
	urldate = {2025-09-13},
	journal = {Monthly Notices of the Royal Astronomical Society},
	author = {Ahrer, Eva-Maria and Gandhi, Siddharth and Alderson, Lili and Kirk, James and Teske, Johanna and Booth, Richard A and McDonald, Catriona H and Christie, Duncan A and Claringbold, Alastair B and Nealon, Rebecca and Panwar, Vatsal and Veras, Dimitri and Wakeford, Hannah R and Wheatley, Peter J and Zamyatina, Maria},
	month = jun,
	year = {2025},
	pages = {2535--2554},
}

@article{murphy_evidence_2024,
	title = {Evidence for morning-to-evening limb asymmetry on the cool low-density exoplanet {WASP}-107 b},
	volume = {8},
	issn = {2397-3366},
	url = {https://www.nature.com/articles/s41550-024-02367-9},
	doi = {10.1038/s41550-024-02367-9},
	language = {en},
	number = {12},
	urldate = {2025-06-09},
	journal = {Nature Astronomy},
	author = {Murphy, Matthew M. and Beatty, Thomas G. and Schlawin, Everett and Bell, Taylor J. and Line, Michael R. and Greene, Thomas P. and Parmentier, Vivien and Rauscher, Emily and Welbanks, Luis and Fortney, Jonathan J. and Rieke, Marcia},
	month = sep,
	year = {2024},
	pages = {1562--1574},
}

@article{lecavelier_des_etangs_rayleigh_2008,
	title = {Rayleigh scattering in the transit spectrum of {HD} 189733b},
	volume = {481},
	issn = {0004-6361, 1432-0746},
	url = {http://www.aanda.org/10.1051/0004-6361:200809388},
	doi = {10.1051/0004-6361:200809388},
	language = {en},
	number = {2},
	urldate = {2022-06-20},
	journal = {Astronomy \& Astrophysics},
	author = {Lecavelier des Etangs, A. and Pont, F. and Vidal-Madjar, A. and Sing, D.},
	month = apr,
	year = {2008},
	pages = {L83--L86},
}

@article{feinstein_early_2023,
	title = {Early {Release} {Science} of the exoplanet {WASP}-39b with {JWST} {NIRISS}},
	volume = {614},
	issn = {0028-0836, 1476-4687},
	url = {https://www.nature.com/articles/s41586-022-05674-1},
	doi = {10.1038/s41586-022-05674-1},
	language = {en},
	number = {7949},
	urldate = {2023-02-17},
	journal = {Nature},
	author = {Feinstein, Adina D. and Radica, Michael and Welbanks, Luis and Murray, Catriona Anne and Ohno, Kazumasa and Coulombe, Louis-Philippe and Espinoza, Néstor and Bean, Jacob L. and Teske, Johanna K. and Benneke, Björn and Line, Michael R. and Rustamkulov, Zafar and Saba, Arianna and Tsiaras, Angelos and Barstow, Joanna K. and Fortney, Jonathan J. and Gao, Peter and Knutson, Heather A. and MacDonald, Ryan J. and Mikal-Evans, Thomas and Rackham, Benjamin V. and Taylor, Jake and Parmentier, Vivien and Batalha, Natalie M. and Berta-Thompson, Zachory K. and Carter, Aarynn L. and Changeat, Quentin and dos Santos, Leonardo A. and Gibson, Neale P. and Goyal, Jayesh M. and Kreidberg, Laura and López-Morales, Mercedes and Lothringer, Joshua D. and Miguel, Yamila and Molaverdikhani, Karan and Moran, Sarah E. and Morello, Giuseppe and Mukherjee, Sagnick and Sing, David K. and Stevenson, Kevin B. and Wakeford, Hannah R. and Ahrer, Eva-Maria and Alam, Munazza K. and Alderson, Lili and Allen, Natalie H. and Batalha, Natasha E. and Bell, Taylor J. and Blecic, Jasmina and Brande, Jonathan and Caceres, Claudio and Casewell, S. L. and Chubb, Katy L. and Crossfield, Ian J. M. and Crouzet, Nicolas and Cubillos, Patricio E. and Decin, Leen and Désert, Jean-Michel and Harrington, Joseph and Heng, Kevin and Henning, Thomas and Iro, Nicolas and Kempton, Eliza M.-R. and Kendrew, Sarah and Kirk, James and Krick, Jessica and Lagage, Pierre-Olivier and Lendl, Monika and Mancini, Luigi and Mansfield, Megan and May, E. M. and Mayne, N. J. and Nikolov, Nikolay K. and Palle, Enric and Petit dit de la Roche, Dominique J. M. and Piaulet, Caroline and Powell, Diana and Redfield, Seth and Rogers, Laura K. and Roman, Michael T. and Roy, Pierre-Alexis and Nixon, Matthew C. and Schlawin, Everett and Tan, Xianyu and Tremblin, P. and Turner, Jake D. and Venot, Olivia and Waalkes, William C. and Wheatley, Peter J. and Zhang, Xi},
	month = jan,
	year = {2023},
	pages = {670--675},
}

@ARTICLE{Kitzmann:18,
       author = {{Kitzmann}, Daniel and {Heng}, Kevin},
        title = "{Optical properties of potential condensates in exoplanetary atmospheres}",
      journal = {\mnras},
         year = 2018,
        month = mar,
       volume = {475},
       number = {1},
        pages = {94-107},
          doi = {10.1093/mnras/stx3141},
archivePrefix = {arXiv},
       eprint = {1710.04946},
 primaryClass = {astro-ph.EP},
       adsurl = {https://ui.adsabs.harvard.edu/abs/2018MNRAS.475...94K}
}

@ARTICLE{Heng:16,
       author = {{Heng}, Kevin},
        title = "{A Cloudiness Index for Transiting Exoplanets Based on the Sodium and Potassium Lines: Tentative Evidence for Hotter Atmospheres Being Less Cloudy at Visible Wavelengths}",
      journal = {\apjl},
         year = 2016,
        month = jul,
       volume = {826},
       number = {1},
          eid = {L16},
        pages = {L16},
          doi = {10.3847/2041-8205/826/1/L16},
archivePrefix = {arXiv},
       eprint = {1606.07218},
 primaryClass = {astro-ph.EP},
       adsurl = {https://ui.adsabs.harvard.edu/abs/2016ApJ...826L..16H}
}

@misc{PLATON7,
      title={PLATON 7: An ultra-fast exoplanet atmospheric retrieval code}, 
      author={Michael Zhang and Tyler Fairnington and Yayaati Chachan and Kimberly Paragas and Eliza Kempton and Heather Knutson and Jacob Bean},
      year={2026},
      eprint={2611.26680},
      archivePrefix={arXiv},
      primaryClass={astro-ph.IM},
      url={https://arxiv.org/abs/2611.26680}, 
}

@article{Mukherjee_2026,
doi = {10.3847/2041-8213/aeb6af},
url = {https://doi.org/10.3847/2041-8213/aeb6af},
year = {2026},
month = {nov},
publisher = {The American Astronomical Society},
volume = {1011},
number = {2},
pages = {L40},
author = {{Mukherjee}, Sagnick and Welbanks, Luis and Fu, Guangwei and Sing, David K. and Line, Michael R.},
title = {A New Method for Constraining 3D Exoplanet Atmospheres: Application to WASP-94A b’s Panchromatic Spectrum},
journal = {The Astrophysical Journal Letters}
}

@INCOLLECTION{Madhu:18,
       author = {{Madhusudhan}, Nikku},
        title = "{Atmospheric Retrieval of Exoplanets}",
    booktitle = {Handbook of Exoplanets},
         year = 2018,
       editor = {{Deeg}, Hans J. and {Belmonte}, Juan Antonio},
          eid = {104},
        pages = {104},
          doi = {10.1007/978-3-319-55333-7_104},
       adsurl = {https://ui.adsabs.harvard.edu/abs/2018haex.bookE.104M}
}

@ARTICLE{Mendoca:18,
       author = {{Mendon{\c{c}}a}, Jo{\~a}o M. and {Tsai}, Shang-min and {Malik}, Matej and {Grimm}, Simon L. and {Heng}, Kevin},
        title = "{Three-dimensional Circulation Driving Chemical Disequilibrium in WASP-43b}",
      journal = {\apj},
         year = 2018,
        month = dec,
       volume = {869},
       number = {2},
          eid = {107},
        pages = {107},
          doi = {10.3847/1538-4357/aaed23},
archivePrefix = {arXiv},
       eprint = {1808.00501},
 primaryClass = {astro-ph.EP},
       adsurl = {https://ui.adsabs.harvard.edu/abs/2018ApJ...869..107M}
}

@ARTICLE{Lines:18,
       author = {{Lines}, S. and {Manners}, J. and {Mayne}, N.~J. and {Goyal}, J. and {Carter}, A.~L. and {Boutle}, I.~A. and {Lee}, Elspeth and {Helling}, Ch and {Drummond}, B. and {Acreman}, D.~M. and {Sing}, D.~K.},
        title = "{Exonephology: transmission spectra from a 3D simulated cloudy atmosphere of HD 209458b}",
      journal = {\mnras},
         year = 2018,
        month = nov,
       volume = {481},
       number = {1},
        pages = {194-205},
          doi = {10.1093/mnras/sty2275},
archivePrefix = {arXiv},
       eprint = {1808.05887},
 primaryClass = {astro-ph.EP},
       adsurl = {https://ui.adsabs.harvard.edu/abs/2018MNRAS.481..194L}
}

@ARTICLE{Mayne:17,
       author = {{Mayne}, Nathan J. and {Debras}, Florian and {Baraffe}, Isabelle and {Thuburn}, John and {Amundsen}, David S. and {Acreman}, David M. and {Smith}, Chris and {Browning}, Matthew K. and {Manners}, James and {Wood}, Nigel},
        title = "{Results from a set of three-dimensional numerical experiments of a hot Jupiter atmosphere}",
      journal = {\aap},
         year = 2017,
        month = aug,
       volume = {604},
          eid = {A79},
        pages = {A79},
          doi = {10.1051/0004-6361/201730465},
archivePrefix = {arXiv},
       eprint = {1704.00539},
 primaryClass = {astro-ph.EP},
       adsurl = {https://ui.adsabs.harvard.edu/abs/2017A&A...604A..79M}
}

@article{zamyatina_quenching-driven_2024,
	title = {Quenching-driven equatorial depletion and limb asymmetries in hot {Jupiter} atmospheres: {WASP}-96b example},
	volume = {529},
	copyright = {https://creativecommons.org/licenses/by/4.0/},
	issn = {0035-8711, 1365-2966},
	shorttitle = {Quenching-driven equatorial depletion and limb asymmetries in hot {Jupiter} atmospheres},
	url = {https://academic.oup.com/mnras/article/529/2/1776/7615449},
	doi = {10.1093/mnras/stae600},
	language = {en},
	number = {2},
	urldate = {2024-04-03},
	journal = {Monthly Notices of the Royal Astronomical Society},
	author = {Zamyatina, Maria and Christie, Duncan A and Hébrard, Eric and Mayne, Nathan J and Radica, Michael and Taylor, Jake and Baskett, Harry and Moore, Ben and Lils, Craig and Sergeev, Denis E and Ahrer, Eva-Maria and Manners, James and Kohary, Krisztian and Feinstein, Adina D},
	month = mar,
	year = {2024},
	pages = {1776--1801},
}

@article{fournier-tondreau_near-infrared_2024,
	title = {Near-infrared transmission spectroscopy of {HAT}-{P}-18 b with {NIRISS}: {Disentangling} planetary and stellar features in the era of \textit{{JWST}}},
	volume = {528},
	issn = {0035-8711, 1365-2966},
	shorttitle = {Near-infrared transmission spectroscopy of {HAT}-{P}-18 b with {NIRISS}},
	url = {https://academic.oup.com/mnras/article/528/2/3354/7468143},
	doi = {10.1093/mnras/stad3813},
	language = {en},
	number = {2},
	urldate = {2024-02-13},
	journal = {Monthly Notices of the Royal Astronomical Society},
	author = {Fournier-Tondreau, Marylou and MacDonald, Ryan J and Radica, Michael and Lafrenière, David and Welbanks, Luis and Piaulet, Caroline and Coulombe, Louis-Philippe and Allart, Romain and Morel, Kim and Artigau, Étienne and Albert, Loïc and Lim, Olivia and Doyon, René and Benneke, Björn and Rowe, Jason F and Darveau-Bernier, Antoine and Cowan, Nicolas B and Lewis, Nikole K and Cook, Neil J and Flagg, Laura and Genest, Frédéric and Pelletier, Stefan and Johnstone, Doug and Dang, Lisa and Kaltenegger, Lisa and Taylor, Jake and Turner, Jake D},
	month = jan,
	year = {2024},
	pages = {3354--3377},
}

@article{radica_muted_2024,
	title = {Muted {Features} in the {JWST} {NIRISS} {Transmission} {Spectrum} of {Hot} {Neptune} {LTT} 9779b},
	volume = {962},
	issn = {2041-8205, 2041-8213},
	url = {https://iopscience.iop.org/article/10.3847/2041-8213/ad20e4},
	doi = {10.3847/2041-8213/ad20e4},
	language = {en},
	number = {1},
	urldate = {2024-02-13},
	journal = {The Astrophysical Journal Letters},
	author = {Radica, Michael and Coulombe, Louis-Philippe and Taylor, Jake and Albert, Loic and Allart, Romain and Benneke, Björn and Cowan, Nicolas B. and Dang, Lisa and Lafrenière, David and Thorngren, Daniel and Artigau, Étienne and Doyon, René and Flagg, Laura and Johnstone, Doug and Pelletier, Stefan and Roy, Pierre-Alexis},
	month = feb,
	year = {2024},
	pages = {L20},
}

@article{radica_awesome_2023,
	title = {Awesome {SOSS}: transmission spectroscopy of {WASP}-96b with {NIRISS}/{SOSS}},
	volume = {524},
	issn = {0035-8711, 1365-2966},
	shorttitle = {Awesome {SOSS}},
	url = {https://academic.oup.com/mnras/article/524/1/835/7198120},
	doi = {10.1093/mnras/stad1762},
	language = {en},
	number = {1},
	urldate = {2023-08-09},
	journal = {Monthly Notices of the Royal Astronomical Society},
	author = {Radica, Michael and Welbanks, Luis and Espinoza, Néstor and Taylor, Jake and Coulombe, Louis-Philippe and Feinstein, Adina D and Goyal, Jayesh and Scarsdale, Nicholas and Albert, Loïc and Baghel, Priyanka and Bean, Jacob L and Blecic, Jasmina and Lafrenière, David and MacDonald, Ryan J and Zamyatina, Maria and Allart1, Romain and Artigau, Étienne and Batalha, Natasha E and Cook, Neil James and Cowan, Nicolas B and Dang, Lisa and Doyon, René and Fournier-Tondreau, Marylou and Johnstone, Doug and Line, Michael R and Moran, Sarah E and Mukherjee, Sagnick and Pelletier, Stefan and Roy, Pierre-Alexis and Talens, Geert Jan and Filippazzo, Joseph and Pontoppidan, Klaus and Volk, Kevin},
	month = jul,
	year = {2023},
	pages = {835--856},
}

@article{tsai_photochemically_2023,
	title = {Photochemically produced {SO2} in the atmosphere of {WASP}-39b},
	volume = {617},
	issn = {0028-0836, 1476-4687},
	url = {https://www.nature.com/articles/s41586-023-05902-2},
	doi = {10.1038/s41586-023-05902-2},
	language = {en},
	number = {7961},
	urldate = {2023-05-25},
	journal = {Nature},
	author = {Tsai, Shang-Min and Lee, Elspeth K. H. and Powell, Diana and Gao, Peter and Zhang, Xi and Moses, Julianne and Hébrard, Eric and Venot, Olivia and Parmentier, Vivien and Jordan, Sean and Hu, Renyu and Alam, Munazza K. and Alderson, Lili and Batalha, Natalie M. and Bean, Jacob L. and Benneke, Björn and Bierson, Carver J. and Brady, Ryan P. and Carone, Ludmila and Carter, Aarynn L. and Chubb, Katy L. and Inglis, Julie and Leconte, Jérémy and Line, Michael and López-Morales, Mercedes and Miguel, Yamila and Molaverdikhani, Karan and Rustamkulov, Zafar and Sing, David K. and Stevenson, Kevin B. and Wakeford, Hannah R. and Yang, Jeehyun and Aggarwal, Keshav and Baeyens, Robin and Barat, Saugata and De Val-Borro, Miguel and Daylan, Tansu and Fortney, Jonathan J. and France, Kevin and Goyal, Jayesh M. and Grant, David and Kirk, James and Kreidberg, Laura and Louca, Amy and Moran, Sarah E. and Mukherjee, Sagnick and Nasedkin, Evert and Ohno, Kazumasa and Rackham, Benjamin V. and Redfield, Seth and Taylor, Jake and Tremblin, Pascal and Visscher, Channon and Wallack, Nicole L. and Welbanks, Luis and Youngblood, Allison and Ahrer, Eva-Maria and Batalha, Natasha E. and Behr, Patrick and Berta-Thompson, Zachory K. and Blecic, Jasmina and Casewell, S. L. and Crossfield, Ian J. M. and Crouzet, Nicolas and Cubillos, Patricio E. and Decin, Leen and Désert, Jean-Michel and Feinstein, Adina D. and Gibson, Neale P. and Harrington, Joseph and Heng, Kevin and Henning, Thomas and Kempton, Eliza M.-R. and Krick, Jessica and Lagage, Pierre-Olivier and Lendl, Monika and Lothringer, Joshua D. and Mansfield, Megan and Mayne, N. J. and Mikal-Evans, Thomas and Palle, Enric and Schlawin, Everett and Shorttle, Oliver and Wheatley, Peter J. and Yurchenko, Sergei N.},
	month = may,
	year = {2023},
	pages = {483--487},
}

@article{roman_clouds_2021,
	title = {Clouds in {Three}-dimensional {Models} of {Hot} {Jupiters} over a {Wide} {Range} of {Temperatures}. {I}. {Thermal} {Structures} and {Broadband} {Phase}-curve {Predictions}},
	volume = {908},
	issn = {0004-637X, 1538-4357},
	url = {https://iopscience.iop.org/article/10.3847/1538-4357/abd549},
	doi = {10.3847/1538-4357/abd549},
	language = {en},
	number = {1},
	urldate = {2022-01-14},
	journal = {The Astrophysical Journal},
	author = {Roman, Michael T. and Kempton, Eliza M.-R. and Rauscher, Emily and Harada, Caleb K. and Bean, Jacob L. and Stevenson, Kevin B.},
	month = feb,
	year = {2021},
	pages = {101},
}

@article{thorngren_connecting_2019,
	title = {Connecting {Giant} {Planet} {Atmosphere} and {Interior} {Modeling}: {Constraints} on {Atmospheric} {Metal} {Enrichment}},
	volume = {874},
	issn = {2041-8213},
	shorttitle = {Connecting {Giant} {Planet} {Atmosphere} and {Interior} {Modeling}},
	url = {https://iopscience.iop.org/article/10.3847/2041-8213/ab1137},
	doi = {10.3847/2041-8213/ab1137},
	language = {en},
	number = {2},
	urldate = {2020-11-02},
	journal = {The Astrophysical Journal},
	author = {Thorngren, Daniel and Fortney, Jonathan J.},
	month = apr,
	year = {2019},
	pages = {L31},
}

@article{espinoza_inhomogeneous_2024,
	title = {Inhomogeneous terminators on the exoplanet {WASP}-39 b},
	volume = {632},
	issn = {0028-0836, 1476-4687},
	url = {https://www.nature.com/articles/s41586-024-07768-4},
	doi = {10.1038/s41586-024-07768-4},
	language = {en},
	number = {8027},
	urldate = {2024-10-09},
	journal = {Nature},
	author = {Espinoza, Néstor and Steinrueck, Maria E. and Kirk, James and MacDonald, Ryan J. and Savel, Arjun B. and Arnold, Kenneth and Kempton, Eliza M.-R. and Murphy, Matthew M. and Carone, Ludmila and Zamyatina, Maria and Lewis, David A. and Samra, Dominic and Kiefer, Sven and Rauscher, Emily and Christie, Duncan and Mayne, Nathan and Helling, Christiane and Rustamkulov, Zafar and Parmentier, Vivien and May, Erin M. and Carter, Aarynn L. and Zhang, Xi and López-Morales, Mercedes and Allen, Natalie and Blecic, Jasmina and Decin, Leen and Mancini, Luigi and Molaverdikhani, Karan and Rackham, Benjamin V. and Palle, Enric and Tsai, Shang-Min and Ahrer, Eva-Maria and Bean, Jacob L. and Crossfield, Ian J. M. and Haegele, David and Hébrard, Eric and Kreidberg, Laura and Powell, Diana and Schneider, Aaron D. and Welbanks, Luis and Wheatley, Peter and Brahm, Rafael and Crouzet, Nicolas},
	month = aug,
	year = {2024},
	pages = {1017--1020},
}

@article{batalha17,
  author =        {{Batalha}, Natasha E. and {Line}, M.~R.},
  journal =       {\aj},
  month =         apr,
  number =        {4},
  pages =         {151},
  title =         {{Information Content Analysis for Selection of
                   Optimal JWST Observing Modes for Transiting Exoplanet
                   Atmospheres}},
  volume =        {153},
  year =          {2017},
  doi =           {10.3847/1538-3881/aa5faa},
  eid =           {151},
}

@article{Ahrer:25a,
  author =        {{Ahrer}, Eva-Maria and {Gandhi}, Siddharth and
                   {Alderson}, Lili and {Kirk}, James and
                   {Teske}, Johanna and {Booth}, Richard A. and
                   {McDonald}, Catriona H. and {Christie}, Duncan A. and
                   {Claringbold}, Alastair B. and {Nealon}, Rebecca and
                   {Panwar}, Vatsal and {Veras}, Dimitri and
                   {Wakeford}, Hannah R. and {Wheatley}, Peter J. and
                   {Zamyatina}, Maria},
  journal =       {\mnras},
  month =         jul,
  number =        {3},
  pages =         {2535-2554},
  title =         {{Tracing the formation and migration history:
                   molecular signatures in the atmosphere of misaligned
                   hot Jupiter WASP-94 A b using JWST NIRSpec/G395H}},
  volume =        {540},
  year =          {2025},
  doi =           {10.1093/mnras/staf819},
}

@article{Radica:24,
  author =        {{Radica}, Michael},
  journal =       {The Journal of Open Source Software},
  month =         aug,
  number =        {100},
  pages =         {6898},
  title =         {{exoTEDRF: An EXOplanet Transit and Eclipse Data
                   Reduction Framework}},
  volume =        {9},
  year =          {2024},
  doi =           {10.21105/joss.06898},
  eid =           {6898},
}

@article{Roman:19,
doi = {10.3847/1538-4357/aafdb5},
url = {https://doi.org/10.3847/1538-4357/aafdb5},
year = {2019},
month = {feb},
publisher = {The American Astronomical Society},
volume = {872},
number = {1},
pages = {1},
author = {Roman, Michael and Rauscher, Emily},
title = {Modeled Temperature-dependent Clouds with Radiative Feedback in Hot Jupiter Atmospheres},
journal = {The Astrophysical Journal}
}

@software{claude5fable,
  author       = {{Anthropic}},
  title        = {Claude (Claude Fable 5 version)},
  year         = {2026},
  version      = {Fable 5},
  publisher    = {Anthropic},
  url          = {https://claude.ai},
  type         = {Large language model}
}

@software{chatgpt56sol,
  author       = {{OpenAI}},
  title        = {ChatGPT (ChatGPT 5.6 Sol version)},
  year         = {2026},
  version      = {5.6 Sol},
  publisher    = {OpenAI},
  url          = {https://chatgpt.com},
  type         = {Large language model}
}

@misc{Savel:26,
      title={Precise Determination of the Metallicity and C/O of WASP-39~b From a Single JWST Instrument Mode with Phase-Resolved Cross-Correlation Retrievals}, 
      author={Arjun B. Savel and Eliza M. -R. Kempton and Erin M. May and Matthew C. Nixon and Jegug Ih and Katherine A. Bennett and Joost P. Wardenier},
      year={2026},
      eprint={2607.18409},
      archivePrefix={arXiv},
      primaryClass={astro-ph.EP},
      url={https://arxiv.org/abs/2607.18409}, 
}

@misc{ACCESS,
author = {Boerner, Timothy J. and Deems, Stephen and Furlani, Thomas R. and Knuth, Shelley L. and Towns, John},
title = {ACCESS: Advancing Innovation: NSF’s Advanced Cyberinfrastructure Coordination Ecosystem: Services \& Support},
year = {2023},
isbn = {9781450399852},
publisher = {Association for Computing Machinery},
address = {New York, NY, USA},
url = {https://doi.org/10.1145/3569951.3597559},
doi = {10.1145/3569951.3597559},
booktitle = {Practice and Experience in Advanced Research Computing 2023: Computing for the Common Good},
pages = {173–176},
numpages = {4},
location = {Portland, OR, USA},
howpublished = {In \emph{Practice and Experience in Advanced Research Computing 2023: Computing for the Common Good}, 173--176 (Association for Computing Machinery, New
  York)},
}

@software{JAX,
  author = {James Bradbury and Roy Frostig and Peter Hawkins and Matthew James Johnson and Yash Katariya and Chris Leary and Dougal Maclaurin and George Necula and Adam Paszke and Jake Vander{P}las and Skye Wanderman-{M}ilne and Qiao Zhang},
  title = {{JAX}: composable transformations of {P}ython+{N}um{P}y programs},
  url = {http://github.com/jax-ml/jax},
  version = {0.3.13},
  year = {2018},
}

@misc{jaxoplanet,
  author =        {{Hattori}, Soichiro and {Garcia}, Lionel and
                   {Murray}, Catriona and {Dong}, Jiayin and
                   {Dholakia}, Shashank and {Degen}, David and
                   {Foreman-Mackey}, Daniel},
  month =         may,
  publisher =     {Zenodo},
  title =         {{exoplanet-dev/jaxoplanet: Astronomical time series
                   analysis with JAX}},
  year =          {2025},
  doi =           {10.5281/zenodo.10736936},
  eid =           {10.5281/zenodo.10736936},
}

@article{power2,
  author =        {{Maxted}, P.~F.~L.},
  journal =       {\aap},
  month =         aug,
  pages =         {A39},
  title =         {{Comparison of the power-2 limb-darkening law from
                   the STAGGER-grid to Kepler light curves of transiting
                   exoplanets}},
  volume =        {616},
  year =          {2018},
  doi =           {10.1051/0004-6361/201832944},
  eid =           {A39},
}

@article{power_2_better,
  author =        {{Morello}, G. and {Tsiaras}, A. and {Howarth}, I.~D. and
                   {Homeier}, D.},
  journal =       {\aj},
  month =         sep,
  number =        {3},
  pages =         {111},
  title =         {{High-precision Stellar Limb-darkening in
                   Exoplanetary Transits}},
  volume =        {154},
  year =          {2017},
  doi =           {10.3847/1538-3881/aa8405},
  eid =           {111},
}

@article{exotic-LD,
  author =        {{Grant}, David and {Wakeford}, Hannah},
  journal =       {The Journal of Open Source Software},
  month =         aug,
  number =        {100},
  pages =         {6816},
  title =         {{ExoTiC-LD: thirty seconds to stellar limb-darkening
                   coefficients}},
  volume =        {9},
  year =          {2024},
  doi =           {10.21105/joss.06816},
  eid =           {6816},
}

@article{stagger,
  author =        {{Magic}, Z. and {Chiavassa}, A. and {Collet}, R. and
                   {Asplund}, M.},
  journal =       {\aap},
  month =         jan,
  pages =         {A90},
  title =         {{The Stagger-grid: A grid of 3D stellar atmosphere
                   models. IV. Limb darkening coefficients}},
  volume =        {573},
  year =          {2015},
  doi =           {10.1051/0004-6361/201423804},
  eid =           {A90},
}

@article{HoffmanGelman:14,
  author =        {{Hoffman}, Matthew D. and {Gelman}, Andrew},
  journal =       {Journal of Machine Learning Research},
  pages =         {1593--1623},
  title =         {{The No-U-Turn Sampler: Adaptively Setting Path
                   Lengths in Hamiltonian Monte Carlo}},
  volume =        {15},
  year =          {2014},
}

@article{numpyro,
  author =        {{Phan}, Du and {Pradhan}, Neeraj and
                   {Jankowiak}, Martin},
  journal =       {arXiv e-prints},
  month =         dec,
  pages =         {arXiv:1912.11554},
  title =         {{Composable Effects for Flexible and Accelerated
                   Probabilistic Programming in NumPyro}},
  year =          {2019},
  doi =           {10.48550/arXiv.1912.11554},
  eid =           {arXiv:1912.11554},
}

@article{Zhang:20,
  author =        {{Zhang}, Michael and {Chachan}, Yayaati and
                   {Kempton}, Eliza M.-R. and {Knutson}, Heather A. and
                   {Chang}, Wenjun (Happy)},
  journal =       {\apj},
  month =         aug,
  number =        {1},
  pages =         {27},
  title =         {{PLATON II: New Capabilities and a Comprehensive
                   Retrieval on HD 189733b Transit and Eclipse Data}},
  volume =        {899},
  year =          {2020},
  doi =           {10.3847/1538-4357/aba1e6},
  eid =           {27},
}

@article{Zhang:25,
  author =        {{Zhang}, Michael and {Paragas}, Kimberly and
                   {Bean}, Jacob L. and {Yeung}, Joseph and
                   {Chachan}, Yayaati and {Greene}, Thomas P. and
                   {Lunine}, Jonathan and {Deming}, Drake},
  journal =       {\aj},
  month =         jan,
  number =        {1},
  pages =         {38},
  title =         {{Retrievals on NIRCam Transmission and Emission
                   Spectra of HD 189733b with PLATON 6, a GPU Code for
                   the JWST Era}},
  volume =        {169},
  year =          {2025},
  doi =           {10.3847/1538-3881/ad8cd2},
  eid =           {38},
}

@article{Fastchem,
  author =        {{Stock}, Joachim W. and {Kitzmann}, Daniel and
                   {Patzer}, A. Beate C.},
  journal =       {\mnras},
  month =         dec,
  number =        {3},
  pages =         {4070-4080},
  title =         {{FASTCHEM 2 : an improved computer program to
                   determine the gas-phase chemical equilibrium
                   composition for arbitrary element distributions}},
  volume =        {517},
  year =          {2022},
  doi =           {10.1093/mnras/stac2623},
}

@article{Asplund:21,
  author =        {{Asplund}, M. and {Amarsi}, A.~M. and {Grevesse}, N.},
  journal =       {\aap},
  month =         sep,
  pages =         {A141},
  title =         {{The chemical make-up of the Sun: A 2020 vision}},
  volume =        {653},
  year =          {2021},
  doi =           {10.1051/0004-6361/202140445},
  eid =           {A141},
}

@article{Polyansky:18,
  author =        {{Polyansky}, Oleg L. and {Kyuberis}, Aleksandra A. and
                   {Zobov}, Nikolai F. and {Tennyson}, Jonathan and
                   {Yurchenko}, Sergei N. and {Lodi}, Lorenzo},
  journal =       {\mnras},
  month =         oct,
  number =        {2},
  pages =         {2597-2608},
  title =         {{ExoMol molecular line lists XXX: a complete
                   high-accuracy line list for water}},
  volume =        {480},
  year =          {2018},
  doi =           {10.1093/mnras/sty1877},
}

@article{Yurchenko:20,
  author =        {{Yurchenko}, S.~N. and {Mellor}, Thomas M. and
                   {Freedman}, Richard S. and {Tennyson}, J.},
  journal =       {\mnras},
  month =         aug,
  number =        {4},
  pages =         {5282-5291},
  title =         {{ExoMol line lists - XXXIX. Ro-vibrational molecular
                   line list for CO$_{2}$}},
  volume =        {496},
  year =          {2020},
  doi =           {10.1093/mnras/staa1874},
}

@article{Li:15,
  author =        {{Li}, Gang and {Gordon}, Iouli E. and
                   {Rothman}, Laurence S. and {Tan}, Yan and
                   {Hu}, Shui-Ming and {Kassi}, Samir and
                   {Campargue}, Alain and {Medvedev}, Emile S.},
  journal =       {\apjs},
  month =         jan,
  number =        {1},
  pages =         {15},
  title =         {{Rovibrational Line Lists for Nine Isotopologues of
                   the CO Molecule in the X
                   $^{1}${\ensuremath{\Sigma}}$^{+}$ Ground Electronic
                   State}},
  volume =        {216},
  year =          {2015},
  doi =           {10.1088/0067-0049/216/1/15},
  eid =           {15},
}

@article{Yurchenko:24,
  author =        {Yurchenko, Sergei N and Owens, Alec and
                   Kefala, Kyriaki and Tennyson, Jonathan},
  journal =       {Monthly Notices of the Royal Astronomical Society},
  month =         {02},
  number =        {2},
  pages =         {3719-3729},
  title =         {ExoMol line lists – LVII. High accuracy
                   ro-vibrational line list for methane (CH4)},
  volume =        {528},
  year =          {2024},
  doi =           {10.1093/mnras/stae148},
  issn =          {0035-8711},
  url =           {https://doi.org/10.1093/mnras/stae148},
}

@article{Allard:19,
  author =        {{Allard}, N.~F. and {Spiegelman}, F. and
                   {Leininger}, T. and {Molliere}, P.},
  journal =       {\aap},
  month =         aug,
  pages =         {A120},
  title =         {{New study of the line profiles of sodium perturbed
                   by H$_{2}$}},
  volume =        {628},
  year =          {2019},
  doi =           {10.1051/0004-6361/201935593},
  eid =           {A120},
}

@article{Allard:16,
  author =        {{Allard}, N.~F. and {Spiegelman}, F. and
                   {Kielkopf}, J.~F.},
  journal =       {\aap},
  month =         may,
  pages =         {A21},
  title =         {{K-H$_{2}$ line shapes for the spectra of cool brown
                   dwarfs}},
  volume =        {589},
  year =          {2016},
  doi =           {10.1051/0004-6361/201628270},
  eid =           {A21},
}

@article{Azzam:16,
  author =        {{Azzam}, Ala'a. A.~A. and {Tennyson}, Jonathan and
                   {Yurchenko}, Sergei N. and {Naumenko}, Olga V.},
  journal =       {\mnras},
  month =         aug,
  number =        {4},
  pages =         {4063-4074},
  title =         {{ExoMol molecular line lists - XVI. The
                   rotation-vibration spectrum of hot H$_{2}$S}},
  volume =        {460},
  year =          {2016},
  doi =           {10.1093/mnras/stw1133},
}

@article{Barber:14,
  author =        {{Barber}, R.~J. and {Strange}, J.~K. and {Hill}, C. and
                   {Polyansky}, O.~L. and {Mellau}, G. Ch. and
                   {Yurchenko}, S.~N. and {Tennyson}, Jonathan},
  journal =       {\mnras},
  month =         jan,
  number =        {2},
  pages =         {1828-1835},
  title =         {{ExoMol line lists - III. An improved hot
                   rotation-vibration line list for HCN and HNC}},
  volume =        {437},
  year =          {2014},
  doi =           {10.1093/mnras/stt2011},
}

@article{Coles:19,
  author =        {{Coles}, Phillip A. and {Yurchenko}, Sergei N. and
                   {Tennyson}, Jonathan},
  journal =       {\mnras},
  month =         dec,
  number =        {4},
  pages =         {4638-4647},
  title =         {{ExoMol molecular line lists - XXXV. A
                   rotation-vibration line list for hot ammonia}},
  volume =        {490},
  year =          {2019},
  doi =           {10.1093/mnras/stz2778},
}

@article{Underwood:16,
  author =        {{Underwood}, Daniel S. and {Tennyson}, Jonathan and
                   {Yurchenko}, Sergei N. and {Huang}, Xinchuan and
                   {Schwenke}, David W. and {Lee}, Timothy J. and
                   {Clausen}, S{\o}nnik and {Fateev}, Alexander},
  journal =       {\mnras},
  month =         jul,
  number =        {4},
  pages =         {3890-3899},
  title =         {{ExoMol molecular line lists - XIV. The
                   rotation-vibration spectrum of hot SO$_{2}$}},
  volume =        {459},
  year =          {2016},
  doi =           {10.1093/mnras/stw849},
}

@article{Guillot:10,
  author =        {{Guillot}, T.},
  journal =       {\aap},
  month =         sep,
  pages =         {A27},
  title =         {{On the radiative equilibrium of irradiated planetary
                   atmospheres}},
  volume =        {520},
  year =          {2010},
  doi =           {10.1051/0004-6361/200913396},
  eid =           {A27},
}

@article{Pymultinest,
  author =        {{Buchner}, J. and {Georgakakis}, A. and {Nandra}, K. and
                   {Hsu}, L. and {Rangel}, C. and {Brightman}, M. and
                   {Merloni}, A. and {Salvato}, M. and {Donley}, J. and
                   {Kocevski}, D.},
  journal =       {\aap},
  month =         apr,
  pages =         {A125},
  title =         {{X-ray spectral modelling of the AGN obscuring region
                   in the CDFS: Bayesian model selection and catalogue}},
  volume =        {564},
  year =          {2014},
  doi =           {10.1051/0004-6361/201322971},
  eid =           {A125},
}

@article{Skilling:2006,
  author =        {John Skilling},
  journal =       {Bayesian Analysis},
  number =        {4},
  pages =         {833 -- 859},
  publisher =     {International Society for Bayesian Analysis},
  title =         {{Nested sampling for general Bayesian computation}},
  volume =        {1},
  year =          {2006},
  doi =           {10.1214/06-BA127},
  url =           {https://doi.org/10.1214/06-BA127},
}

@article{Feroz:09,
  author =        {{Feroz}, Farhan and {Hobson}, M.~P. and
                   {Bridges}, Mike},
  journal =       {\mnras},
  pages =         {1601--1614},
  title =         {{MultiNest: An Efficient and Robust Bayesian
                   Inference Tool for Cosmology and Particle Physics}},
  volume =        {398},
  year =          {2009},
  doi =           {10.1111/j.1365-2966.2009.14548.x},
}

@article{Kokori:23,
  author =        {{Kokori}, A. and others},
  journal =       {\apjs},
  number =        {1},
  pages =         {4},
  title =         {{ExoClock Project. III. 450 New Exoplanet Ephemerides
                   from Ground- and Space-based Observations}},
  volume =        {265},
  year =          {2023},
  doi =           {10.3847/1538-4365/ac9da4},
}

@article{Atreya:22,
  author =        {{Atreya}, Sushil K. and {Crida}, Aur{\'e}lien and
                   {Guillot}, Tristan and {Li}, Cheng and {Lunine},
                   Jonathan I. and {Madhusudhan}, Nikku and {Mousis},
                   Olivier and {Wong}, Michael H.},
  title =         {{The Origin and Evolution of Saturn: A Post-Cassini
                   Perspective}},
  journal =       {arXiv e-prints},
  year =          {2022},
  eid =           {arXiv:2205.06914},
  doi =           {10.48550/arXiv.2205.06914},
  eprint =        {2205.06914},
  archiveprefix = {arXiv},
  primaryclass =  {astro-ph.EP},
}

@article{Sousa:21,
  author =        {{Sousa}, S.~G. and {Adibekyan}, V. and
                   {Delgado-Mena}, E. and {Santos}, N.~C. and
                   {Rojas-Ayala}, B. and {Soares}, B.~M.~T.~B. and
                   {Legoinha}, H. and {Ulmer-Moll}, S. and others},
  title =         {{SWEET-Cat 2.0: The Cat Just Got SWEETer. Higher
                   Quality Spectra and Precise Parallaxes from Gaia eDR3}},
  journal =       {\aap},
  volume =        {656},
  pages =         {A53},
  year =          {2021},
  doi =           {10.1051/0004-6361/202141584},
}

@article{Parmentier:16,
  author =        {{Parmentier}, Vivien and {Fortney}, Jonathan J. and {Showman}, Adam P. and {Morley}, Caroline V. and {Marley}, Mark S.},
  journal =       {\apj},
  pages =         {22},
  title =         {{Transitions in the Cloud Composition of Hot Jupiters}},
  volume =        {828},
  year =          {2016},
  doi =           {10.3847/0004-637X/828/1/22},
  eid =           {22},
}

@article{Lines:19,
  author =        {{Lines}, S. and {Mayne}, N. J. and {Manners}, J. and {Boutle}, I. A. and {Drummond}, B. and {Mikal-Evans}, T. and {Kohary}, K. and {Sing}, D. K.},
  journal =       {\mnras},
  pages =         {1332--1355},
  title =         {{Overcast on Osiris: 3D Radiative-hydrodynamical Simulations of a Cloudy Hot Jupiter Using the Parametrized, Phase-equilibrium Cloud Formation Code EddySed}},
  volume =        {488},
  year =          {2019},
  doi =           {10.1093/mnras/stz1788},
}

@article{Rackham:18,
  author =        {{Rackham}, Benjamin V. and {Apai}, D{\'a}niel and {Giampapa}, Mark S.},
  journal =       {\apj},
  pages =         {122},
  title =         {{The Transit Light Source Effect: False Spectral Features and Incorrect Densities for M-dwarf Transiting Planets}},
  volume =        {853},
  year =          {2018},
  doi =           {10.3847/1538-4357/aaa08c},
  eid =           {122},
}

@article{Molliere:19,
  author = {{Molli\`ere}, P. and {Wardenier}, J.~P. and {van Boekel}, R. and
            {Henning}, Th. and {Molaverdikhani}, K. and {Snellen}, I.~A.~G.},
  title = {{petitRADTRANS: A Python radiative transfer package for exoplanet characterization and retrieval}},
  journal = {\aap},
  year = {2019},
  volume = {627},
  pages = {A67},
  doi = {10.1051/0004-6361/201935470}
}

@Article{Hunter:2007,
    Author = {Hunter, J. D.},
    Title = {Matplotlib: A 2D graphics environment},
    Journal = {Computing in Science \& Engineering},
    Volume = {9},
    Number = {3},
    Pages = {90--95},
    publisher = {IEEE COMPUTER SOC},
    doi = {10.1109/MCSE.2007.55},
    year = 2007
}

@misc{bushouse_2023,
    author = {Bushouse, Howard and Eisenhamer, Jonathan and Dencheva, Nadia and Davies, James and  Greenfield, Perry and Morrison, Jane and Hodge, Phil and Simon, Bernie and Grumm, David and Droettboom, Michael and Slavich, Edward and Sosey, Megan and Pauly, Tyler and Miller, Todd and Jedrzejewski, Robert and Hack, Warren and Davis, David and Crawford, Steven and Law, David and Gordon, Karl and Regan, Michael and Cara, Mihai and MacDonald, Ken and Bradley, Larry and Shanahan, Clare and Jamieson, William and Teodoro, Mairan and Williams, Thomas and Pena-Guerrero, Maria},
    title = {JWST Calibration Pipeline},
    month = oct,
    year = 2023,
    note = {{If you use this software in your work, please cite it using the following metadata.}},
    publisher = {Zenodo},
    version = {1.12.3},
    doi = {10.5281/zenodo.8404029},
    url = {https://doi.org/10.5281/zenodo.8404029}
}

@Article{harris2020array,
    title = {Array programming with {NumPy}},
    author = {Charles R. Harris and K. Jarrod Millman and St{\'{e}}fan J. van der Walt and Ralf Gommers and Pauli Virtanen and David Cournapeau and Eric Wieser and Julian Taylor and Sebastian Berg and Nathaniel J. Smith and Robert Kern and Matti Picus and Stephan Hoyer and Marten H. van Kerkwijk and Matthew Brett and Allan Haldane and Jaime Fern{\'{a}}ndez del R{\'{i}}o and Mark Wiebe and Pearu Peterson and Pierre G{\'{e}}rard-Marchant and Kevin Sheppard and Tyler Reddy and Warren Weckesser and Hameer Abbasi and Christoph Gohlke and Travis E. Oliphant},
    year = {2020},
    month = sep,
    journal = {Nature},
    volume = {585},
    number = {7825},
    pages = {357--362},
    doi = {10.1038/s41586-020-2649-2},
    publisher = {Springer Science and Business Media {LLC}},
    url = {https://doi.org/10.1038/s41586-020-2649-2}
}

@article{astropy:2013,
    Adsurl = {http://adsabs.harvard.edu/abs/2013A%26A...558A..33A},
    Archiveprefix = {arXiv},
    Author = {{Astropy Collaboration} and {Robitaille}, T.~P. and {Tollerud}, E.~J. and {Greenfield}, P. and {Droettboom}, M. and {Bray}, E. and {Aldcroft}, T. and {Davis}, M. and {Ginsburg}, A. and {Price-Whelan}, A.~M. and {Kerzendorf}, W.~E. and {Conley}, A. and {Crighton}, N. and {Barbary}, K. and {Muna}, D. and {Ferguson}, H. and {Grollier}, F. and {Parikh}, M.~M. and {Nair}, P.~H. and {Unther}, H.~M. and {Deil}, C. and {Woillez}, J. and {Conseil}, S. and {Kramer}, R. and {Turner}, J.~E.~H. and {Singer}, L. and {Fox}, R. and {Weaver}, B.~A. and {Zabalza}, V. and {Edwards}, Z.~I. and {Azalee Bostroem}, K. and {Burke}, D.~J. and {Casey}, A.~R. and {Crawford}, S.~M. and {Dencheva}, N. and {Ely}, J. and {Jenness}, T. and {Labrie}, K. and {Lim}, P.~L. and {Pierfederici}, F. and {Pontzen}, A. and {Ptak}, A. and {Refsdal}, B. and {Servillat}, M. and {Streicher}, O.},
    Doi = {10.1051/0004-6361/201322068},
    Eid = {A33},
    Eprint = {1307.6212},
    Journal = {\aap},
    Month = oct,
    Pages = {A33},
    Primaryclass = {astro-ph.IM},
    Title = {{Astropy: A community Python package for astronomy}},
    Volume = 558,
    Year = 2013
}

@ARTICLE{2020SciPy-NMeth,
    author  = {Virtanen, Pauli and Gommers, Ralf and Oliphant, Travis E. and Haberland, Matt and Reddy, Tyler and Cournapeau, David and Burovski, Evgeni and Peterson, Pearu and Weckesser, Warren and Bright, Jonathan and {van der Walt}, St{\'e}fan J. and Brett, Matthew and Wilson, Joshua and Millman, K. Jarrod and Mayorov, Nikolay and Nelson, Andrew R. J. and Jones, Eric and Kern, Robert and Larson, Eric and Carey, C J and Polat, {\.I}lhan and Feng, Yu and Moore, Eric W. and {VanderPlas}, Jake and Laxalde, Denis and Perktold, Josef and Cimrman, Robert and Henriksen, Ian and Quintero, E. A. and Harris, Charles R. and Archibald, Anne M. and Ribeiro, Ant{\^o}nio H. and Pedregosa, Fabian and {van Mulbregt}, Paul and {SciPy 1.0 Contributors}},
    title   = {{{SciPy} 1.0: Fundamental Algorithms for Scientific Computing in Python}},
    journal = {Nature Methods},
    year    = {2020},
    volume  = {17},
    pages   = {261--272},
    adsurl  = {https://rdcu.be/b08Wh},
    doi     = {10.1038/s41592-019-0686-2},
}

% ========= Appendices =========
\appendix
\restartappendixnumbering
\makeatletter
\@addtoreset{figure}{section}
\@addtoreset{table}{section}
\makeatother

\section{Cross-code radiative-transfer check}\label{app:prt}

We repeated the full-spectrum retrievals with a modified branch of the \texttt{petitRADTRANS} (\texttt{pRT}) radiative transfer code \citep{Molliere:19, pRT3, pRT4}. This check uses the same limb-averaged spectrum, 1- and 2-TP structures, equilibrium-chemistry assumption, and fitted species, as well as the same 1--layer pressure grid ($10^{-9}$--$10^{3}$\,bar), 1-bar reference radius, and $1/r^2$ gravity. We use the line-by-line opacity cross-sections at $R=20{,}000$ to match the \platon\ implementation. Both codes compute the transit radius from slant optical depths through spherical shells, treating scattering as pure extinction. They differ in their equilibrium-chemistry solver, opacity tables, and aerosol--opacity parameterizations, and in three numerical details: the hydrostatic integration, the quadrature of the annulus area, and how the opaque cloud deck is placed on the pressure grid. The comparison is therefore a test of robustness across codes. The \texttt{pRT} solution gives $\zsun=\prtlinearz$, compared with $\twodlinearz$ from \platon; both are far below their respective 1-TP values of $\onedlinearz$ for \texttt{PLATON} and $\prtonedlinearz$ for \texttt{pRT} (Figure~\ref{fig:prtcheck}). The best fit has $\chi^2=\prtchisq$, modestly worse than $\chi^2=\twodchisq$ for \platon. The cross-code calculation thus supports the the conclusions of this paper preferring the low metallicity solution.

\begin{figure*}[h!]
\centering
\includegraphics[width=\textwidth]{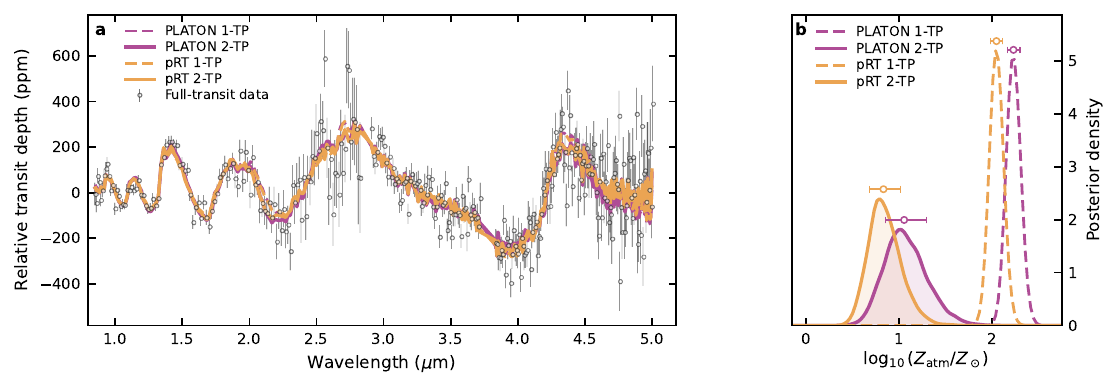}
\caption{Cross-code robustness check. Panel (a) compares the best-fitting 2-TP transmission spectra from \platon\ and \texttt{pRT} for the same retained limb-averaged data. Panel (b) compares their atmospheric-metallicity posteriors with the 1-TP \platon\ and \texttt{pRT} posteriors shown as the dashed curves. Both 2-TP radiative-transfer calculations strongly prefer the low metallicity of the 2-TP solution.}
\label{fig:prtcheck}
\end{figure*}
% =====
\section{Numerical validation of \texttt{PLATON 7.0}}
\label{app:platonvalidation}

We evaluate the legacy \platon\ 6.0 and 32-bit \texttt{JAX} calculations at identical parameter states from the fiducial WASP-94Ab two-component retrieval. At the saved best fit, the two spectra differ by an RMS of \platonsevenrmsppm\ ppm and a maximum of \platonsevenmaxppm\ ppm across the 316 wavelength bins. The largest discrepancy is $\platonsevenmaxsigma$ of the measurement uncertainty in the same bin (Figure~\ref{fig:platonvalidation}). 

We test the inference itself by selecting \platonsevenvalidationdraws\ evenly spaced draws from the equal-weight \platon\ 7.0 posterior and reevaluating every draw with the legacy likelihood. All selected states are valid in both implementations. We assign each draw an importance weight proportional to $\exp(\ln\mathcal{L}_{6.0}-\ln\mathcal{L}_{7.0})$, yielding the metallicity posterior expected from the legacy calculation without repeating the nested-sampling run. The reweighted posterior retains an effective-sample fraction within $\platonsevenessfraction$ of 1 and shifts the metallicity mean by only $\platonsevenmetalshift$ posterior standard deviations. Numerical precision therefore has a negligible effect on either the spectrum or the inferred metallicity.

\begin{figure*}[h!]
\centering
\includegraphics[width=0.82\textwidth]{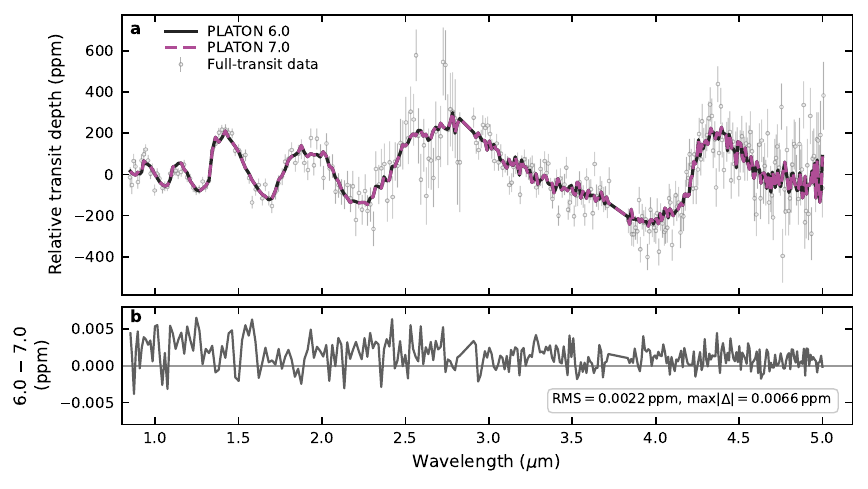}
\caption{Numerical validation of \platon\ 7.0 for the fiducial WASP-94Ab two-component retrieval. The upper panel compares the measured limb-averaged spectrum with the \platon\ 6.0 and 7.0 calculations evaluated at identical saved best-fitting parameters. The lower panel shows their wavelength-dependent difference and reports its RMS and maximum absolute value.}
\label{fig:platonvalidation}
\end{figure*}

\section{Joint posterior distributions}

\begin{figure*}[ht!]
\centering
\begin{minipage}[b]{0.41\textwidth}
\centering
    \includegraphics[width=\linewidth, keepaspectratio]{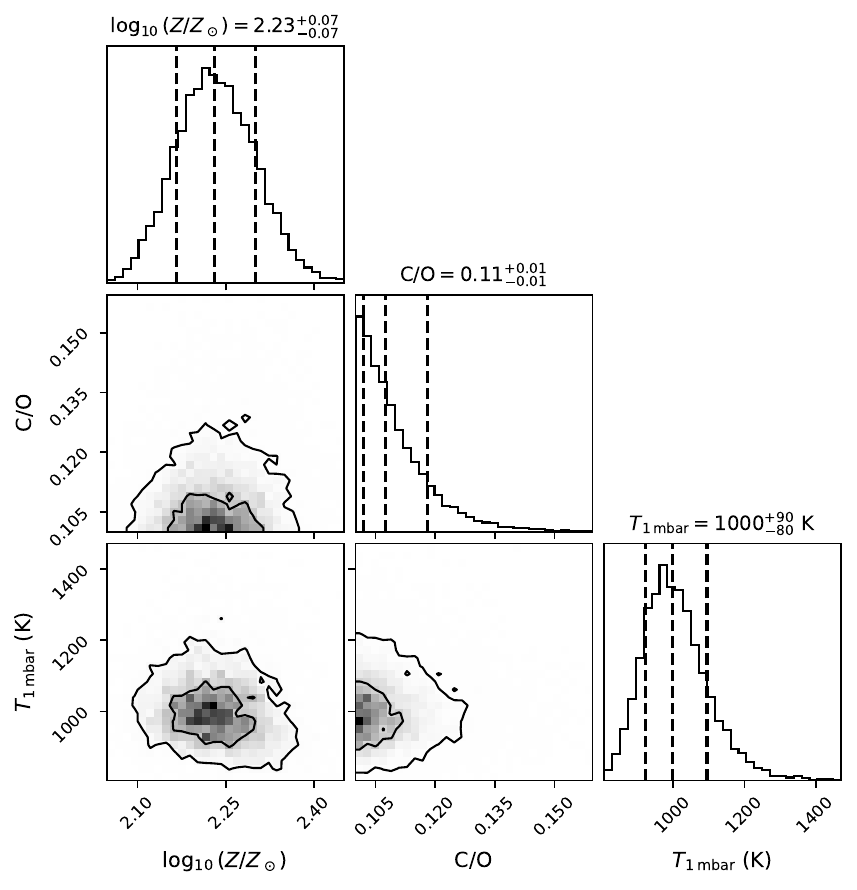}
    \caption{Posteriors of the metallicity, C/O, and temperature for the fiducial 1-TP retrieval. The temperature is evaluated at 0.1 mbar from the Guillot profile of each posterior sample. The panel titles provide the median and 68\% credible intervals, with dashed lines marking the 16th, 50th and 84th percentiles, and the contours enclosing 39.3\% and 86.5\% probabilities.}
    \label{app:corner1D}
\end{minipage}\hfill
\begin{minipage}[b]{0.54\textwidth}
    \centering
    \includegraphics[width=\linewidth, keepaspectratio]{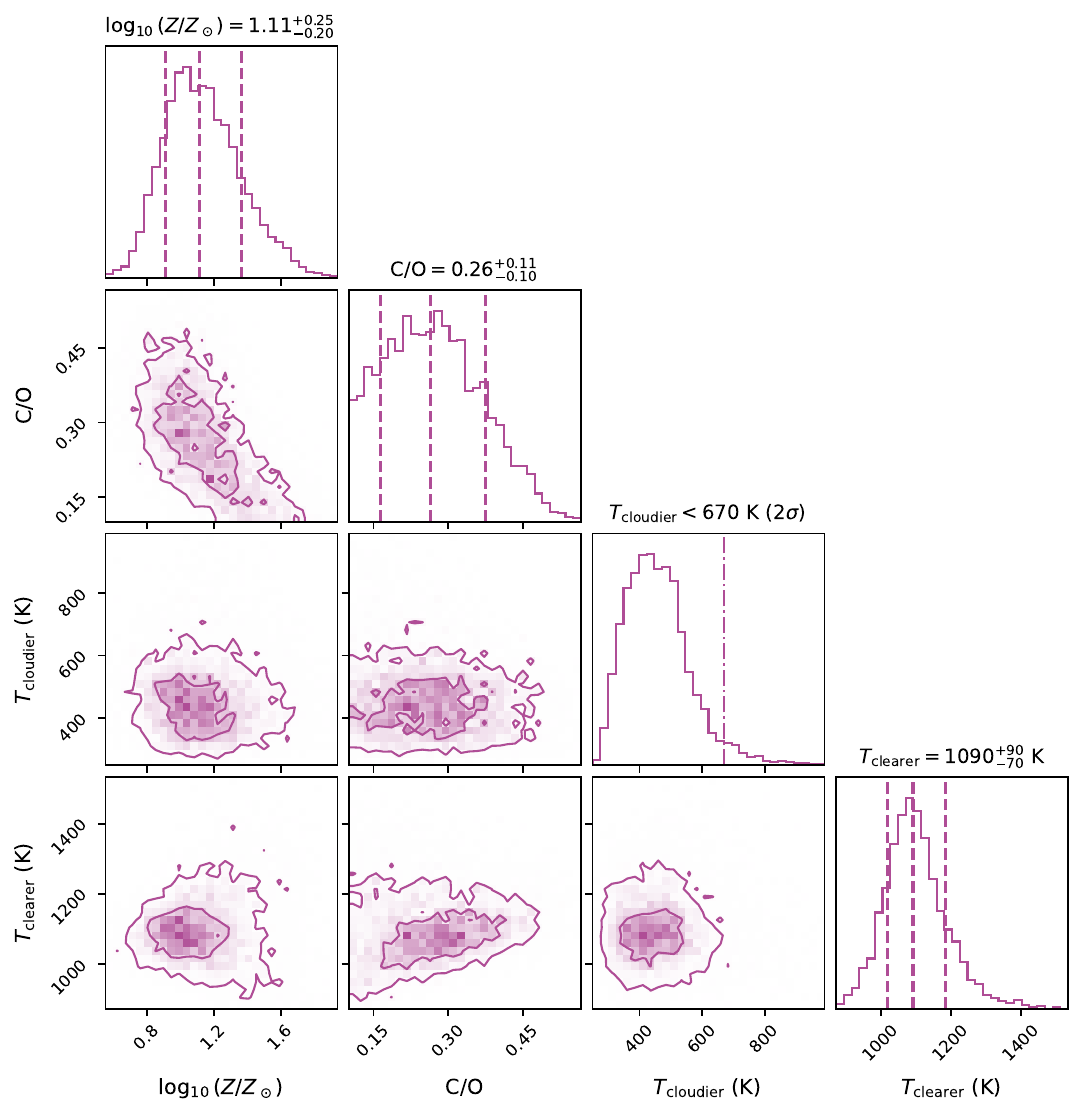}
    \caption{Same as Figure~\ref{app:corner1D} but for the 2-TP retrieval.}
    \label{app:corner1.5D}
\end{minipage}
\end{figure*}

Figures~\ref{app:corner1D} and \ref{app:corner1.5D} show the posteriors of the metallicity, C/O, and temperature for the fiducial 1-TP and 2-TP retrieval. Because the full posteriors have high dimensionality, we provide the full plots of every fitted parameter as vector PDFs in the Zenodo repository (\url{https://zenodo.org/records/23247451}).

\section{Sensitivity tests}
\label{app:D}

\begin{table*}[t]
\centering
\caption{Sensitivity Tests}
\label{tab:sensitivity}
\footnotesize
\begin{tabular}{lcccccc}
\toprule
2-TP retrieval & $\Delta\ln\mathcal{Z}$ & $\log_{10}(\zsun)$ & C/O & $T_{\rm hot}$ (K) & $T_{\rm cold}$ (K) & $f_{\rm cold}$ \\
\midrule
Guillot, shared $\kappa_{\rm th}$ (fiducial) & 0 & $\twodlogz$ & $0.26^{+0.11}_{-0.10}$ & $1090^{+90}_{-70}$ & $<670$ & $\limbweight$ \\
Guillot, separate $\kappa_{\rm th}$ & $+0.17\pm0.25$ & $1.12^{+0.26}_{-0.22}$ & $0.25^{+0.11}_{-0.09}$ & $1080^{+100}_{-80}$ & $<710$ & $0.49^{+0.05}_{-0.05}$ \\
Isothermal & $-0.79\pm0.25$ & $0.98^{+0.29}_{-0.20}$ & $0.29^{+0.10}_{-0.10}$ & $1110^{+90}_{-60}$ & $<780$ & $0.47^{+0.05}_{-0.05}$ \\
\bottomrule
\end{tabular}
\par\smallskip
\parbox{0.98\textwidth}{\footnotesize
\textit{Note.} Values are posterior medians with 68\% credible intervals, except for $T_{\rm cold}$, which is the $2\sigma$ upper limit. Temperatures are evaluated at 1\,mbar, and $\Delta\ln\mathcal{Z}$ is relative to the fiducial retrieval.}
\end{table*}

We test two choices in the TP structure of the 2-TP model: sharing the thermal opacity between the components, and the Guillot profile itself. Table~\ref{tab:sensitivity} compares these tests with the fiducial results.

$T_{\rm int}$ should naturally be shared as it sets the deep atmosphere common to both limbs, whereas sharing $\kappa_{\rm th}$ is less clearly justified. We performed the analysis with a per-limb $\kappa_{\rm th}$. The cloudier component's opacity spans nearly its entire prior ($-1.9<\log_{10}\kappa_{\rm th}<0.3$ at $2\sigma$, against a prior of $-2$ to $0.5$), and the clearer component's is only weakly constrained ($-1.9$ to $0$), as is when it is shared ($-1.9$ to $-0.2$). The metallicity, C/O, temperatures of each limb and other parameters are all unchanged, and the evidence does not favor the additional parameter, so we retain a single shared $\kappa_{\rm th}$. 

Another question is whether the choice in temperature profile drastically changes our inference. To this extent, we also tested isothermal profiles, which gives a slightly warmer clear component temperature ($T_{\rm hot}=1110_{-60}^{+90}\,K$) and the metallicity slightly lower ($[\rm{M/H}] = 0.98_{-0.20}^{+0.29}$), both within $1\sigma$ of the fiducial values. The Guillot model is only preferred by $\Delta \ln{\mathcal{Z}}\approx1$, but we retain it as the fiducial model since it ties both limbs to a common deep atmosphere and can include a quench pressure where CH$_4$ is depleted. The 1-TP retrieval is also insensitive to the profile, with an isothermal giving $[\rm{M/H}]=2.23\pm0.06$ and C/O$=0.11\pm0.01$, and the isothermal 2-TP model heavily favored. The metallicity bias therefore does not depend on the choice of profile.  

The quantity most sensitive to these choices is the cloudier component's temperature upper limit, which goes from $T_{\rm cold}<670$\,K in the fiducial run to $<710$\,K with separate opacities and $<780$\,K with isothermal profiles ($2\sigma$). This is expected given the limit is set by model assumptions (Section~\ref{ssec:coldcomponent}).

\section{Cloudier-component temperature}
\label{app:E}
Figure~\ref{appfig:E1} compares the posterior of the cloudier component's 1-mbar temperature with its prior, and shows how the haze and radius parameters move with that temperature in the fiducial 2-TP posterior (Section~\ref{ssec:coldcomponent}). 

\begin{figure}[h!]
\centering 
\includegraphics[width=0.49\textwidth]{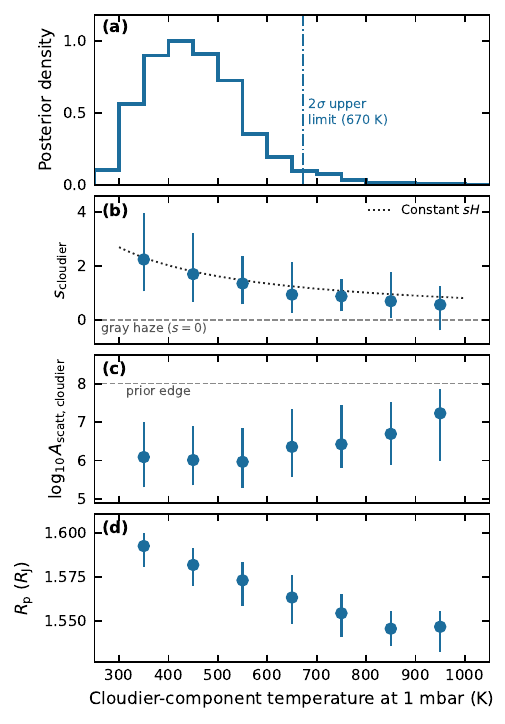}
\caption{The cloudier component's temperature in the fiducial 2-TP retrieval. (a) Posterior of $T_{\rm cold}$ at 1mbar (solid) with the $2\sigma$ upper limit (dash-dotted). (b--d) Posterior median and $1\sigma$ intervals of $s_{\rm cold}$, $\log_{10}A_{\rm scatt,cold}$ and $R_p$ in 100\,K bins of $T_{\rm cold}$. The dotted curve in (b) keeps $sH$ fixed ($s\propto1/T$); dashed lines mark a gray haze in (b) and the upper bound of the amplitude prior in (c).}
\label{appfig:E1}
\end{figure}

\end{document}